\documentclass{article}
\usepackage[top=2.5cm,bottom=2cm,left=2.5cm,right=2.5cm]{geometry}
\usepackage[utf8x]{inputenc}
\usepackage{booktabs}
\usepackage{ctable}
\usepackage{subfigure}
 \usepackage{amsmath}
 \usepackage{natbib}

\begin{document}


\title{Causality in the shock wave/turbulent boundary layer interaction}

\author{Kenzo Sasaki$^1$, Diogo C. Barros$^2$, André V. G. Cavalieri$^1$, Lionel Larchev\^eque$^2$}

\maketitle
             
\begin{abstract}
The mechanisms governing the low-frequency unsteadiness in the shock wave/turbulent boundary layer interaction at Mach 2 are considered.
The investigation is conducted based on the numerical database issued from large-eddy simulations covering approximately 300 cycles of the low-frequency shock fluctuations. 
The evaluation of the spectrum in the interaction zone indicates that the broadband low-frequency unsteadiness is predominantly two-dimensional, and can be isolated via spanwise averaging. 
Empirically derived transfer functions are computed using the averaged flow field, and indicate the occurrence
of a feedback mechanism between downstream flow regions and shock fluctuations. 
The transfer functions are also used as an estimation tool to predict the shock motion accurately; for the largest streamwise separation between input and output signals, correlations above 0.6 are observed between predicted and LES data. 
Computation of spectral proper orthogonal decomposition (SPOD) modes confirms the existence of upstream traveling waves in the leading spectral mode. 
Finally, the spectral modes obtained using selected flow regions downstream of the shock enable the reconstruction of a significant portion of the energy in the interaction zone. 
The current results shed further light on the physical mechanisms driving the shock motion, pointing towards a causal behavior between downstream areas and the characteristic unsteady fluctuations at the approximate shock position.
\end{abstract}

\section{Introduction}

Interactions between shock waves and turbulent flows appear in multiple high-speed applications including supersonic flight and combustion processes \citep{Andreopoulos2000}.
Specifically, in shock wave boundary layer interactions (SBLI), the imposed adverse pressure gradient over the wall induces flow separation, which is associated with a reflected shock presenting highly unsteady low-frequency fluctuations.
This situation occurs in many practical configurations such as turbomachinery flows, transonic buffeting, overexpanded rocket nozzles and supersonic inlet isolators. 
Besides affecting the flow downstream of the interaction zone, these oscillations are responsible for large pressure levels and severe structural loading.
The source of these broadband shock fluctuations has been the subject of extensive debate and comprehensive reviews over the past decades \citep{Delery1985,Dolling2001,Dussauge2006,Babinsky2011,Clemens2014,Gaitonde2015}.

Two mechanisms are recognized to induce reflected shock unsteadiness: upstream boundary-layer forcing and downstream feedback \citep{Clemens2014}.
Among numerous investigations \citep{Plotkin1975, Brusniak1994, Beresh2002}, a strong correlation between the upstream boundary layer streamwise velocity and shock motion was found in the experiments of \citet{Ganapathisubramani2009}. 
Three-dimensional measurements from \citet{Humble2009} demonstrated that low and high-momentum, streamwise-elongated zones convected within the incoming boundary layer affect directly the reflected shock structure.
On the other hand, there is compelling evidence for the influence of the downstream separated flow on the shock motion.
The large-eddy simulations (LES) and linear stability analysis of \citet{Touber2009} suggested the existence of unstable modes and upstream traveling waves associated with a global instability mechanism to be responsible for the low-frequency shock oscillations.
These oscillations were found to be highly correlated to the reattachment pressure measurements performed by \citet{Dupont2006}.
The direct numerical simulation (DNS) of \citet{Wu2008} also pointed to a strong correlation where the shock motion lags the reattachment fluctuations.
The mass-entrainment model proposed by \citet{Piponniau2009} scales the low-frequency content with the separated flow region and highlitghts the importance of the downstream flow on selecting the shock unsteadiness. It is worth noting that a low-frequency feedback from the reattachment region up to the vicinity of the reflected shock was documented for transitional SBLI \cite{sansica_sandham_hu_2016,larchevequeaiaa}. Although the flow features listed above could suggest that a similar mechanism may also be at play for turbulent SBLI, this seems not to have been demonstrated in the current literature.

The low-frequency unsteadiness and the associated mechanisms discussed above were observed for impinging oblique shocks and compression ramp configurations, in both numerical and experimental investigations.
Equally important is the spatial organization of the low-frequency motion in the vicinity of the reflected shock and separation bubble.
The low-pass filtered velocity fields from the simulations of \citet{Priebe2012} highlighted a strong quasi two-dimensional behavior of the reflected shock motion using spanwise-averaging.
The dynamic mode decomposition analysis (DMD) conducted by \citet{Priebe2016} shows that the shock unsteadiness was reproduced using a number of low-frequency DMD modes.
Their spatial imprint qualitatively agreed with the linear unstable modes found in \citet{Touber2009}.
These modes exhibit streamwise-elongated structures in the downstream separation bubble of the compression ramp.
Similar streamwise Görtler-like vortices were found in the DMD analysis from \citet{Pasquariello2017} for a strong impinging shock wave. 
The global stability analysis and DMD computed over spanwise-averaged velocity fields from \citet{Nichols2016} shows a striking resemblance with the recirculation bubble breathing during one cycle of the oscillation.

In the present work, we exploit the LES database from \citet{Jiang2017} where an impinging shock wave interacts with a turbulent boundary layer at Mach number $M=2$ and $Re_\theta\simeq5000$.
This database spans a time domain an order of magnitude larger than the current existing DNS or LES simulations investigated for this configuration \citep{Pirozzoli2006,Touber2009,Pasquariello2017}, corresponding roughly to more than 300 low-frequency cycles.
First, we present a causal analysis in the framework of linear transfer functions to identify the link between the low-frequency shock motion, upstream and downstream mechanisms.
To shed further light on the spatial structure of the low-frequency unsteadiness, we perform a spectral proper orthogonal decomposition (SPOD) of the spanwise-averaged velocity fields in the interaction zone. 
To the authors' knowledge, little is known about this optimal spatial-spectral data decomposition for SBLI configurations.
In section II, a description of the numerical database is provided accounting for the whole set of simulation and flow parameters.
The key spectral features of the shock motion are analyzed in section III.
The mathematical framework to compute the linear transfer functions, causal analysis and spectral decomposition modes is presented in section IV.
We report the prediction and causality results in section V, and the spectral eigenvalues and modes in section VI.
Finally, the concluding remarks are described in section VII.

\section{Numerical data base}

The current large-eddy simulation (LES) reproduces the shock wave/turbulent boundary layer interaction investigated in the experiments of \citet{Schreyer2015}.
The full description of the numerical simulation was presented in \citet{Jiang2017}, together with a thorough validation against wind-tunnel measurements. 
Here, we provide a brief overview of the configuration and main flow parameters.

\begin{figure}
\centering
\includegraphics[scale=0.65]{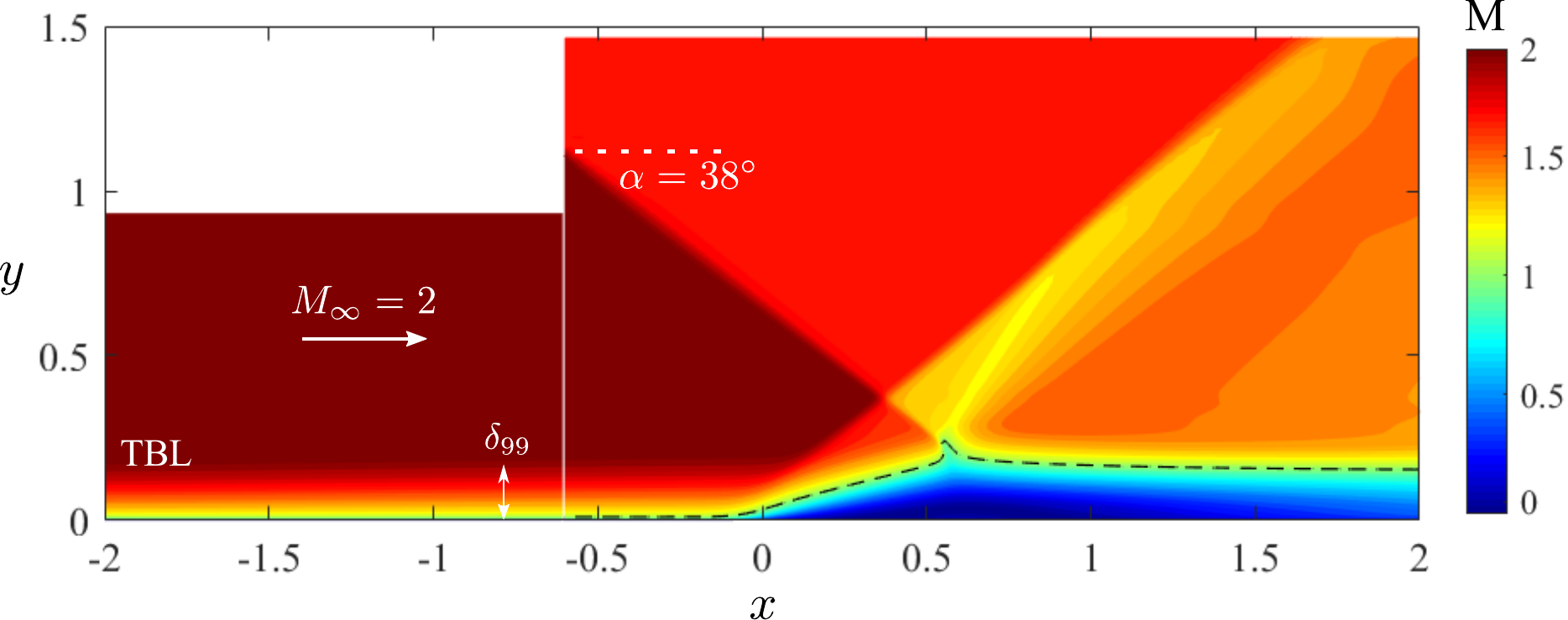}
\caption{Mean Mach number in the streamwise ($x$) wall-normal ($y$) longitudinal plane at $z=0$. The turbulent boundary layer (TBL) thickness ($\delta_{99}$) is shown in the first region of the simulation, and the sonic line is represented in the interaction zone.}
\label{fig:Mean_Mach}
\end{figure}

The numerical setup mimics the experimental shock generated by a flow deflection of $8.5^\circ$ impinging on a $M_{\infty}=2$ turbulent boundary layer with $Re_\theta\simeq 5000$.
The complete set of flow parameters are listed in table~\ref{tab:parameters}. 
The incident shock corresponds to a pressure ratio of 1.58 with an angle of approximately $38^\circ$, as indicated in figure~\ref{fig:Mean_Mach}. 
All simulations were performed using the FLU3M codes from ONERA. 
The code relies on a finite volume discretization in space and an implicit time integration, both being second--order accurate. 
The space scheme is designed to minimize the numerical dissipation by adding the dissipative part of the Roe scheme to a centered scheme in regions with strong compressibility/low vorticity, as identified by means of Ducros'sensor \citep{Roe1981,Ducros1999}. 
The time integration is performed with a maximum CFL number of 11, using 7 sub-iterations to solve the non-linear system in order to ensure that residuals are reduced by at least 1.5 order of magnitude.
The LES modeling is built from an implicit grid filtering coupled with an explicit subgrid modeling through the selective mixed-scale model that was successfully used in previous studies of shock wave/boundary layer interactions \citep{Garnier2002,Agostini2015}.

\setlength{\tabcolsep}{11pt}

\begin{table}
\begin{center}
\caption{Flow parameters.} 
\begin{tabular}{cccccccc}
\\
$M_{\infty}$ & $p_o$ & $T_o$ & $U_{\infty}$ & $\delta_{99}$ & $Re_{\theta}$ & $L$ & $W$   \\ 
\hline \\
2 & $40500\,\text{Pa}$ & $295\,\text{K}$ & $510\,\text{ms}^{-1}$ & $11\,\text{mm}$ & 4850 & $57\,\text{mm}$ & $33\,\text{mm}$ \\  
\hline   
\end{tabular}
\label{tab:parameters}
\end{center}
\end{table}

The wall is modeled as adiabatic. 
Non-reflective boundary conditions relying on characteristic variation in space are set at the inflow, outflow and upper boundaries, while periodicity is used in the spanwise $z$ direction. 
The incident shock is created by enforcing the pre/post-- shock states on the upper boundaries, following \citet{Garnier2002}. 
A fully turbulent inflow condition is obtained by adding stochastic velocity fluctuations to the mean profiles by means of a Synthetic Eddy Method~(SEM) \citep{Jarrin2006}. 
Temperature and density fluctuations are obtained from the velocity fluctuations assuming strong Reynolds analogy and linearised ideal gas law with negligible pressure fluctuations \citep{Agostini2012}. 
The space and time scales required by SEM are obtained from a pre-existing LES of the same $M=2$ boundary layer without shock impingement that encompasses the laminar-to-turbulent transition process \citep{Schreyer2016}.  
Comparisons between the two LES demonstrate that a relaxing length of about 10 boundary layer thicknesses $\delta_{99}$ is sufficient to achieve fully turbulent canonical first-- and second--order statistics from the inflow boundary condition.

The reference mesh was designed by setting the inflow location $16\delta_{99}$ upstream of the shock impingement location. 
The grid resolution and spanwise extent $W$ of the computational domain were increased with respect to previous studies \citep{Agostini2015}.
The corresponding resolution in wall units were $\Delta x^+=28$~(streamwise), $\Delta y_{min}^+=0.85$~(wall--normal) and $\Delta z^+=12$~(spanwise) for $W=3\delta_{99}$. 
Two additional meshes were tested for a grid-convergence study: a refined one with resolution increased by $40\%$, $20\%$ and $30\%$ in the streamwise, wall--normal and spanwise directions, respectively, and an enlarged spanwise domain corresponding to $W=6\delta_{99}$.

A key characteristic of the numerical database is that data is sampled from the reference mesh over approximately 300 periods of the typical low-frequency unsteadiness, as will be discussed later. 
It allows the computation of statistical and spectral quantities, and even conditional statistics with a very low level of statistical uncertainty.
Computations from the two additional meshes were performed for 10 to 18 low-frequency cycles.
These tests were long enough to achieve statistical convergence for validation purposes.

A complete description with validation results from the experimental database and grid convergence can be found in \citet{Jiang2017}.
All the computations and experiments have an interaction length of $L = 5.2\delta_{99}$ and the mean velocity profiles upstream of and within the interaction region agree very well.
Downstream of the interaction region, the LES velocity fields relax toward a canonical boundary layer profile slightly more slowly than in the experiments. 
This was explained by the finite width of the experimental setup with side walls compared to the periodicity set in the computations.

Second-order velocity statistics also show very good agreement between the three computations and the particle image velocimetry (PIV) measurements, with the exception of the streamwise fluctuations in the incoming boundary layer and the initial stage of the mixing layer developing over the separated region. 
In the latter, the LES underestimate the $\overline{u'u'}$ level by 40\% whatever the mesh used. 
Comparisons between the streamwise velocity spectra obtained from the LES and the experiments show that the underestimation originates solely in the low--frequency band $f\,\delta_{99}/U_\infty<0.15$. 
It is therefore postulated that it was due to the intrinsic inability of the present SEM method to generate large-scale convective structures found in the experiments, possibly originating in the nozzle upstream of the flat surface over which the boundary layer develops. 
The otherwise very good agreement between LES and experiments up to the spectral quantities in multiple locations indicate that for the present flow parameters such fluctuations have very limited impact on the dynamics of the interaction.

\section{Spectral analysis of the shock motion}

A space-time spectral analysis is considered to identify the key features of the shock motion. 
The variables are normalized by the interaction length scale $L$ and the free-stream velocity $U_{\infty}$. 
The dimensionless time is defined as $t=t^* U_{\infty}/L$, while the normalized frequency reads $St=fL/{U_{\infty}}$, corresponding to the Strouhal number. 

\begin{figure}
\begin{center}
\includegraphics[scale=0.2]{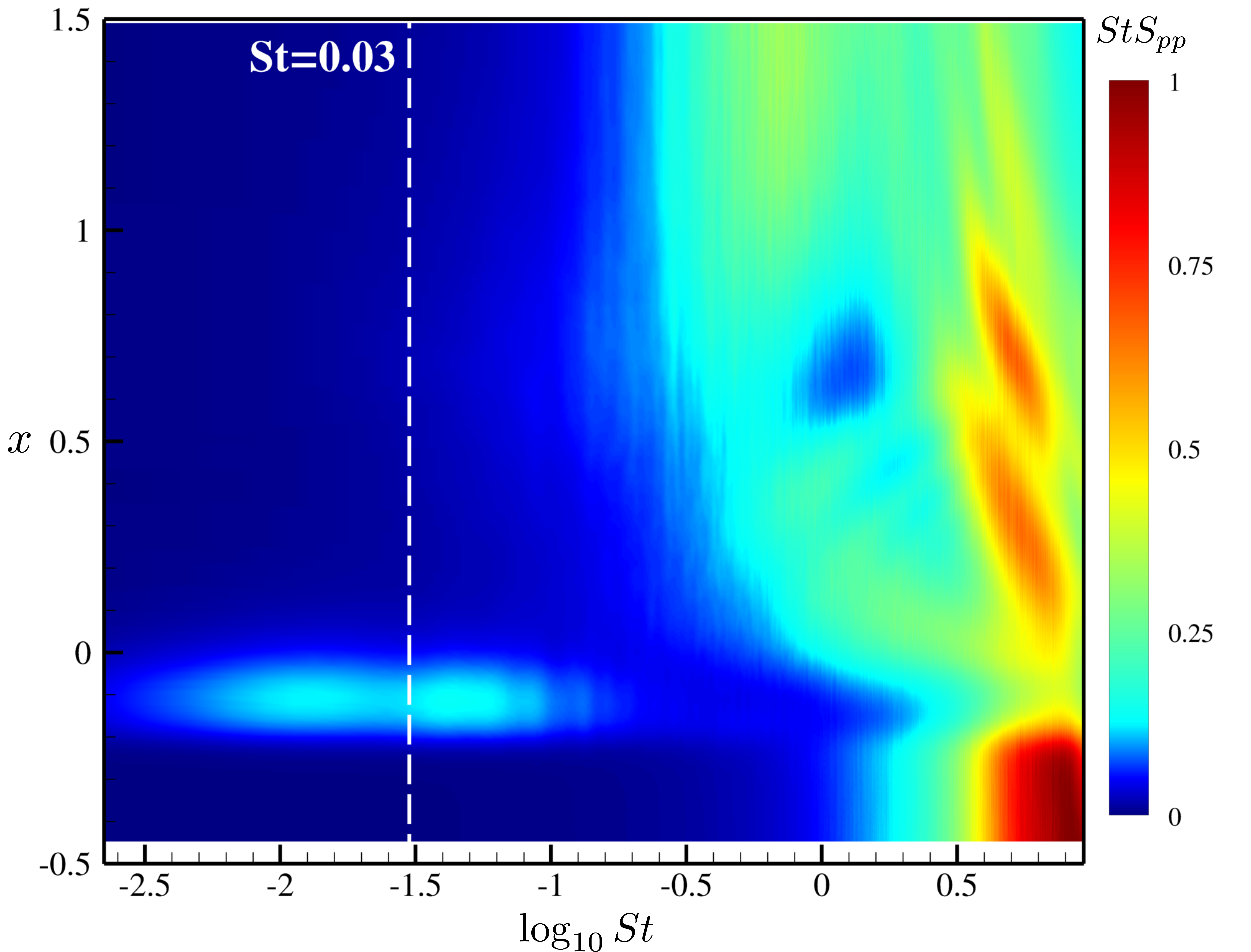}
\end{center}
   \caption{Streamwise evolution of the pre-multiplied and normalized spectrum of the wall pressure fluctuations $p$ ($z=0$).}
  \label{Spectra}
\end{figure}

First, we assess the dominant frequency in the vicinity of the shock. 
The streamwise evolution of the pre-multiplied spectrum of pressure fluctuations is shown in figure \ref{Spectra}.
The spectra is defined as $S_{pp}=\langle \hat{p}\hat{p}^{*}\rangle$, where $\hat{.}$ indicates the Fourier transform from time to frequency $f$, and $\langle{.}\rangle$ defines an ensemble averaging.
In the upstream boundary layer ($x<-0.25$), the spectrum presents mainly high-frequency content ($St>1$) linked to the incoming turbulent eddies. 
However, in the vicinity of the shock position ($-0.2<x<0$), a low-frequency broadband range emerges, and is centered approximately at $St=0.03$.  
The large gap between these frequency scales was reported in previous investigations \citep{Touber2011,Dupont2006}. 
As discussed later in the manuscript, a low-pass filter will be applied to the data to isolate these shock oscillations.

\begin{figure}
\begin{center}
\subfigure[]{\includegraphics[width=0.45\textwidth]{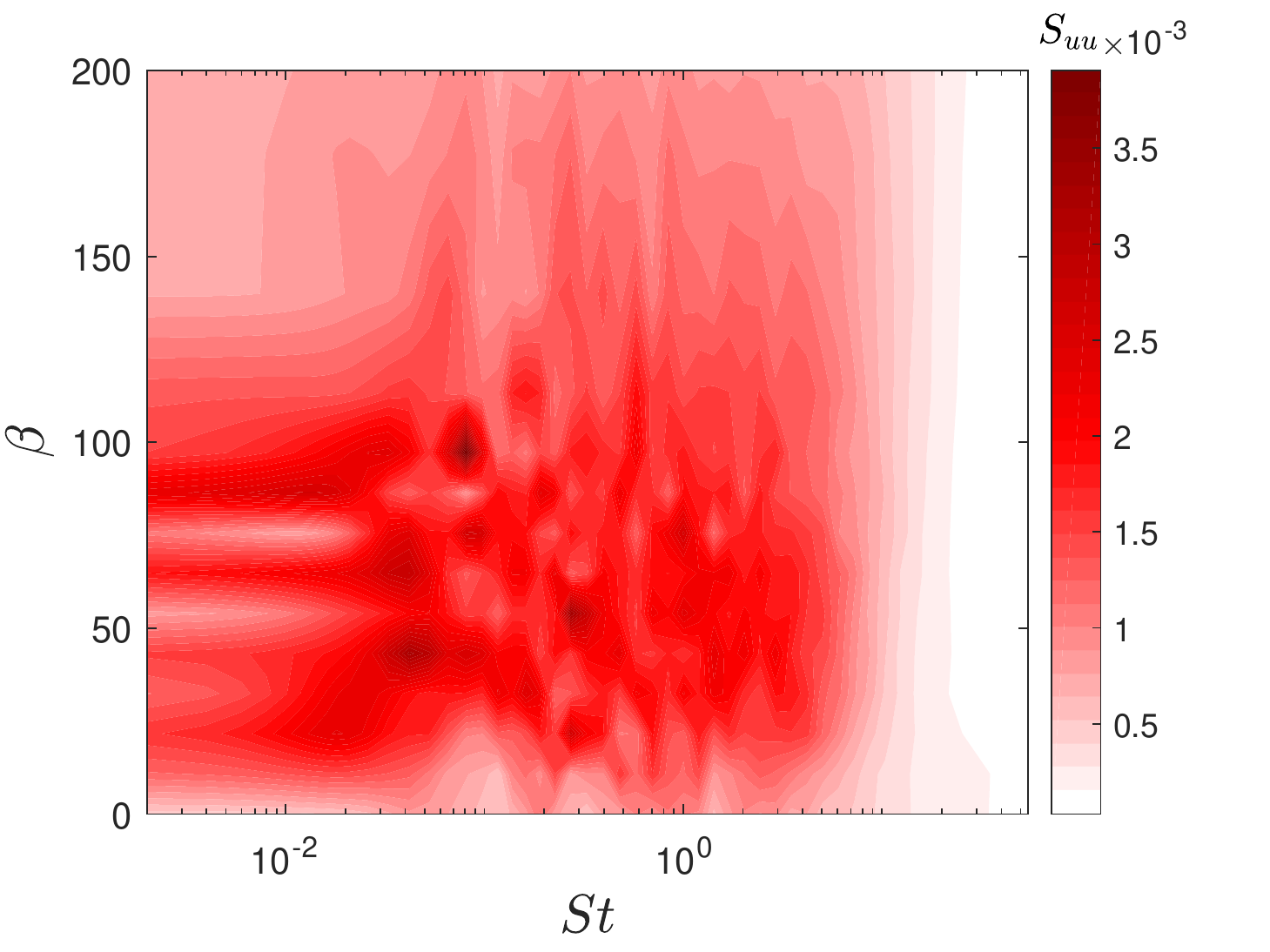}}
\subfigure[]{\includegraphics[width=0.45\textwidth]{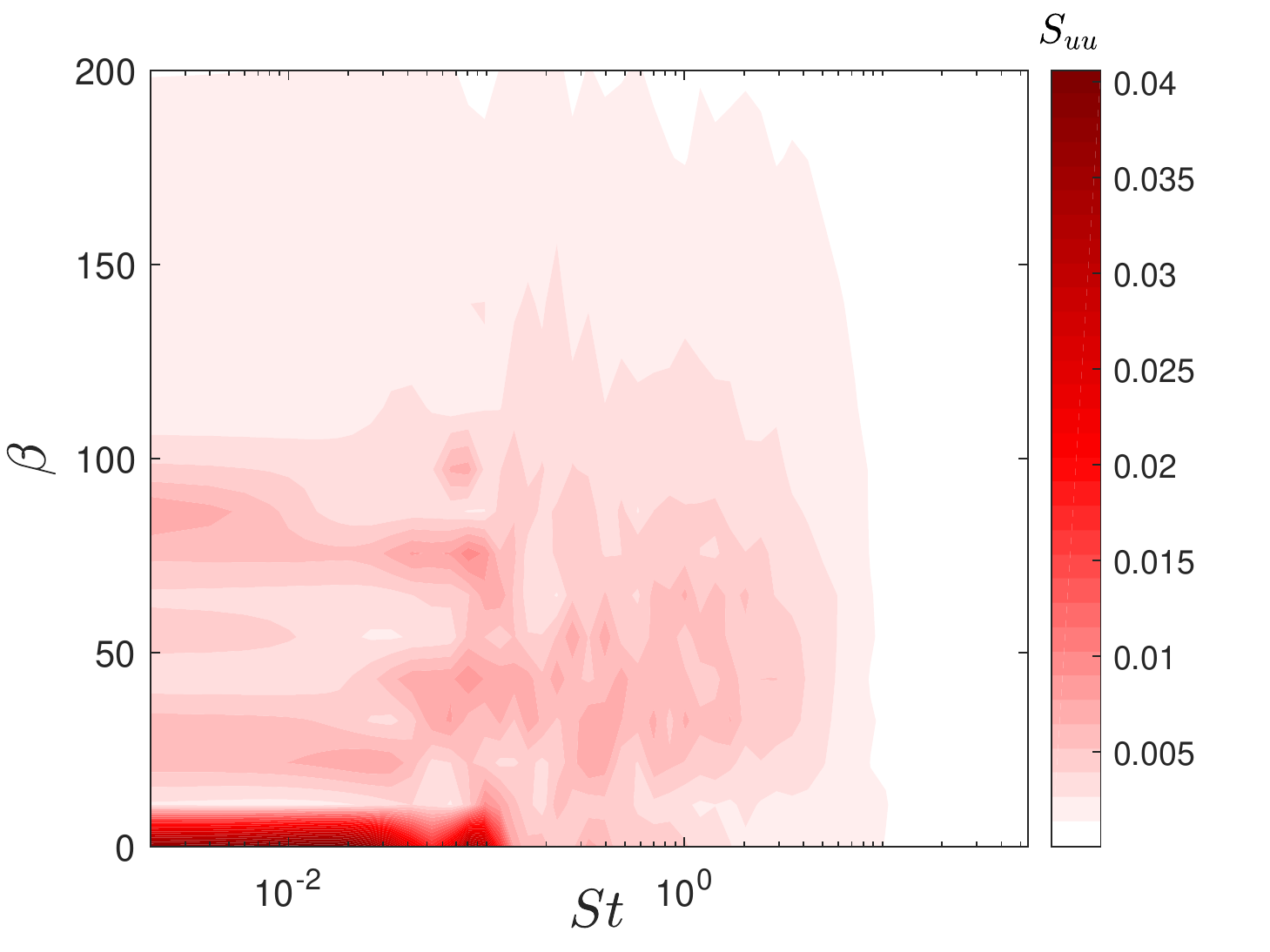}}\\
\subfigure[]{\includegraphics[width=0.45\textwidth]{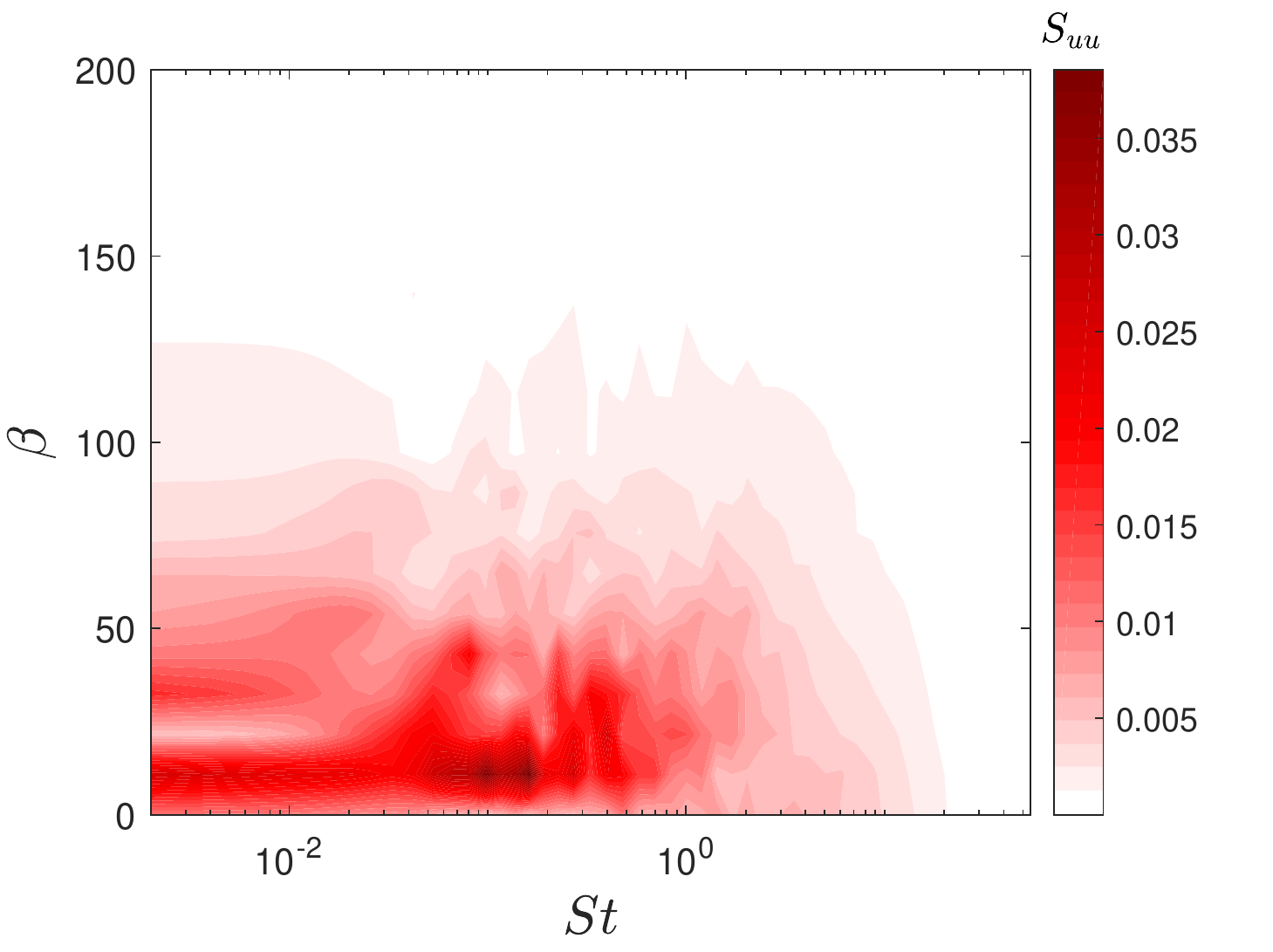}}
\subfigure[]{\includegraphics[width=0.45\textwidth]{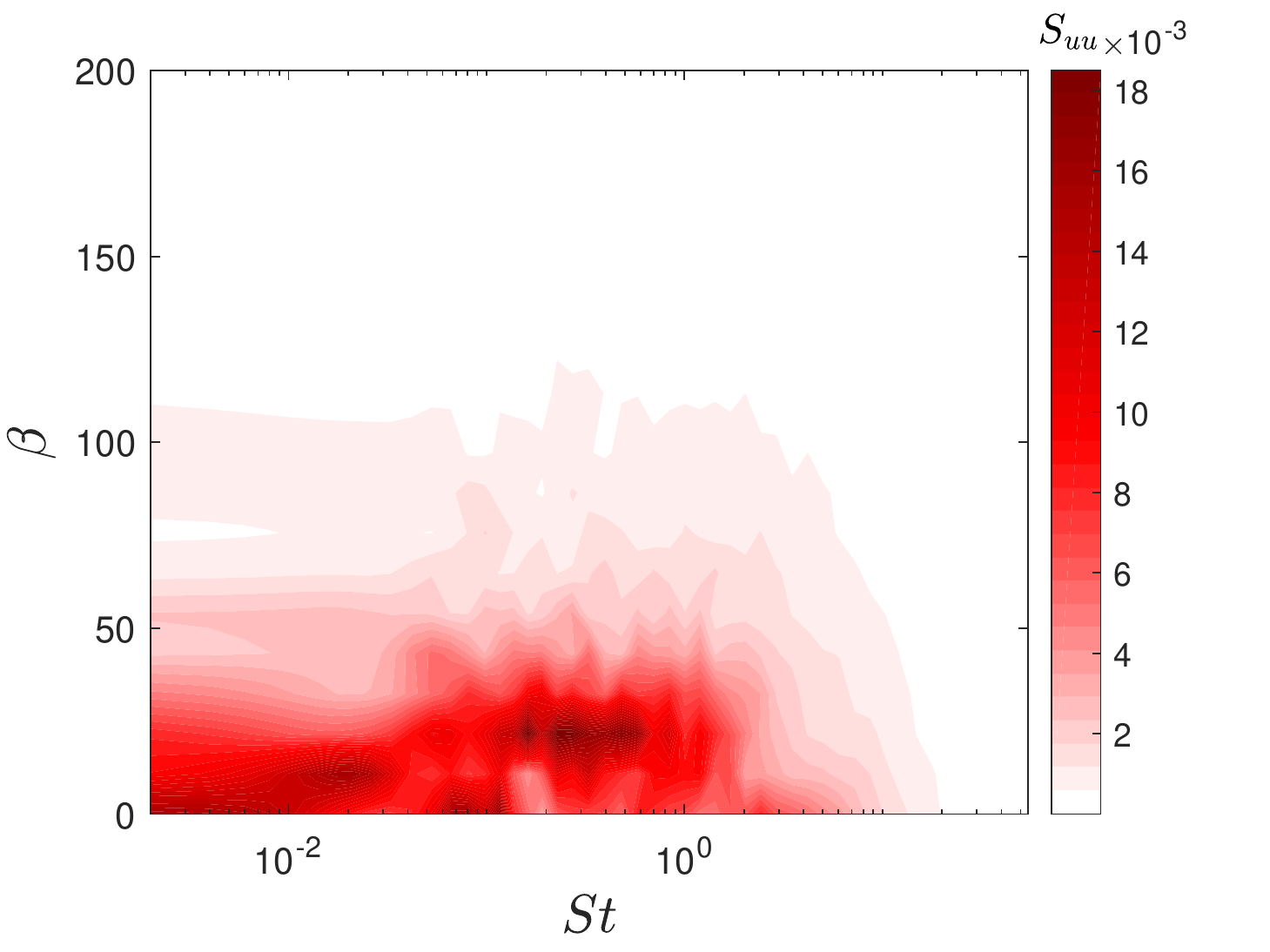}} \\
\end{center}
   \caption{Two-dimensional spectra at $x=-0.35$, 0.0, 0.2 and 1.4, respectively in (a) to (d).}
  \label{2DSpectra}
\end{figure}

To shed further light on the spatial structure of these oscillations, the space-time spectrum is computed for the streamwise velocity fluctuations. 
Here, $S_{uu}(St,\beta)=\langle \hat{\hat{u}}\hat{\hat{u}}^{*}\rangle$, where the double hat indicates a double Fourier transform, from time to $f$ and from $z$ to the spanwise normalized wavenumber $\beta$. 
Figure \ref{2DSpectra} presents the two-dimensional spectra in the near-wall region at four streamwise locations $x = -0.35$, 0.0, 0.2 and 1.4.
These locations correspond to the upstream boundary layer, the approximate shock position, the recirculating bubble and downstream reattachement, respectively.
Upstream of the shock, the flow is dominated by broadband turbulent near-wall streaks. 
In the approximate shock position, the fluctuations are quasi two-dimensional ($\beta=0$).
As discussed above, this region is predominantly governed by low-frequency fluctuations, which can therefore be isolated via spanwise averaging.  
This property will be considered and analyzed in our flow predictions.
The values of $\beta$ increase in the recirculating flow region and further downstream.
The velocity fluctuations exhibit mostly a three-dimensional behavior, and the characteristic frequencies are also higher in this region.
The simultaneous increase of both $St$ and $\beta$ is related to the development of shear-layer vortices and their break-up into 3D eddies.
This behavior can be also attributed to the existence of G\"ortler-type structures developing along the boundary-layer \cite{loginov_adams_zheltovodov_2006,Pirozzoli2006}.

\section{Methods}

\subsection{Linear transfer function: single-input single-output analysis}

A linear, frequency-domain identification method is applied to relate an input signal $I(t)$ to an output $O(t)$ separated in the streamwise/wall-normal plane.
The supporting assumption for the method is the existence of a linear function that maps the input to the output locations.
A similar approach was applied to estimate velocity fluctuations in turbulent jets and boundary layers \citep{sasaki2017real,sasaki2019transfer}.
Our goal is to identify the causal mechanisms behind the low-frequency shock unsteadiness. 
The problem is formulated in the frequency domain,
where the optimal frequency response reads

\begin{equation}
\label{tfequation1}
G_{IO}(f)=\frac{S_{IO}(f)}{S_{II}(f)},
\end{equation}

\noindent where $S_{II}(f)$ and $S_{IO}(f)$ are the auto and cross spectra of the input and output signals, respectively, and $G_{IO}(f)$ refers to the empirical transfer function computed in the frequency domain \citep{bendat2011random}.
These quantities were calculated from the expected values of $\hat{I}(f)\hat{I}^{*}(f)$ and $\hat{I}(f)\hat{O}^{*}(f)$, obtained from an ensemble averaging, via the Welch's method. The spectra was obtained using 230 blocks of 2000 elements each with an overlap of 75\%, which results in a frequency resolution of 0.015. Similar results were obtained using up to 5000 elements in each block.

Equation \eqref{tfequation1} is referred to as the $H_1$ estimator of the system, and it minimizes the error due to measurement noise in the output \cite{rocklin1985comparison}. 
Other formulations, such as the $H_2$ or $H_{\nu}$ estimators, exhibit different performances in terms of sensor noise minimization. 
However, they are expected to perform equally well for this type of estimation, which does not consider the presence of measurement uncertainties and is based in simulation data. 
The $H_1$ estimator presents the interesting property of leading to a prediction error which is linearly uncorrelated with the input \cite{bendat1993spectral,rocklin1985comparison}, therefore resulting in the best linear prediction of the output signal. 

Once the transfer function $G_{IO}(f)$ is computed, its time-domain counterpart is recovered by means of an inverse Fourier transform: 

\begin{equation}
\label{inversefourierfortheestimation}
g_{IO}(t)=\int_{-\infty}^{+\infty} G_{IO}(f) e^{-i 2 \pi f t} df.
\end{equation}

The function $g_{IO}(t)$ represents a linear convolution kernel, and therefore allows one to estimate the output via the convolution with the input signal

\begin{equation}
\label{convolutionforprediction}
O(t)=\int_{-\infty}^{\infty}g_{IO}(\tau)I(t-\tau)\rm d\tau.
\end{equation}

The dummy variable $\tau$ was introduced for the calculation of the convolution. 

\subsection{Hillbert transform to evaluate causality}
\label{hilbert}

The convolution in equation \ref{convolutionforprediction} is taken between $\pm\infty$. 
However, for a causal prediction, this operation should be limited from 0 to $+\infty$, so that only past information is used to predict the output signal. 
The convolution operation is usually limited to the causal part of the convolution kernel, which permits an on-line prediction of the output. 
This type of approach is necessary for active closed-loop applications, for example. 
Nevertheless, the full convolution operation may also be used for data-reconstruction, a feature which was explored in \cite{sasaki2019transfer} for a turbulent boundary layer. 

This implies that if $g_{IO}(t<0) \neq 0$, there is some reverse causality between input and output signals. 
An evaluation of the convolution kernel in the time domain can therefore be used to determine the causality relation between input and output positions. 
Such evaluations in the time domain could be cumbersome and, in some cases, qualitative, particularly for positions which are close to each other, where a feedback path usually occurs. 
To avoid this problem, the evaluation of causality can be performed directly in the frequency domain by means of the Hilbert transform \citep{press2007numerical}. 
This method has the advantage of testing several locations in a fast, computationally efficient manner. 
The Hilbert transform is a linear operation that performs a convolution with the function $1/(\pi f)$,

\begin{equation}
\label{hilberneles}
H(F(f))=\frac{1}{\pi}\int_{-\infty}^{+\infty}\frac{F(\Omega)}{f-\Omega}\mathrm{d}\Omega.
\end{equation}

Hilbert transforms are commonly used in signal processing for the design of causal filters, and the idea was adopted in flow control applications to determine suitable positions for the active control of shear layers \citep{sasaki2018_inpress}. 
A causal filter results when $g_{IO}(t \leq 0)=0$. 
This occurs in convectively unstable flows, such as a Tollmien-Schlichting or Kelvin-Helmholtz dominated shear flows, where upstream measurements are taken to estimate downstream fluctuations. Computation of the Hilbert transform, in this case, is made directly in the frequency-domain by means of a convolution with the previously calculated empirical transfer function.

The approach consists in exploring the following property of the transform: if the imaginary part of the frequency response (interpreted here as a transfer function) of a linear system is equal to the Hilbert transform of the real part, the system represents a causal, linear filter.
Hence, knowledge of the real part is sufficient to completely specify the system, with the imaginary part adding redundant information \cite{bendat2011random}.

Therefore, a quantitative evaluation of the causality of the transfer function $G_{IO}(f)$ may be obtained by computing the correlation between its imaginary part and the Hilbert transform of its real part. This parameter is given by 

\begin{equation}
\label{pparamter}
P=\frac{\int_{-\infty}^{\infty}H(\Re[G_{IO}(f)])\Im[G_{IO}(f)]df}{\sqrt{\int_{-\infty}^{\infty}H(\Re[G_{IO}(f)])^2df}\sqrt{\int_{-\infty}^{\infty}\Im[G_{IO}(f)]^2df}}.
\end{equation}

Values of $P$ close to unity indicate a causal behaviour between input and output positions, with a reverse causality occurring as $P$ approaches zero. 
This parameter will be used in the following section to evaluate the causality between two different locations.

\subsection{Spectral proper orthogonal decomposition}

In the current work, spectral proper orthogonal decomposition (SPOD) is applied in the streamwise/wall-normal plane.
As outlined in section III, the low-frequency shock dynamics can be considered mostly two-dimensional. 
Therefore, spanwise averaging of the fourteen streamwise/wall-normal planes available from the LES data set is considered to isolate the $\beta=0$ fluctuations.  
The effect of the spanwise averaging on the signal prediction framework discussed above is reported in section V (figure \ref{examplecorrelationsfasdf}).

The objective is to extract spatially-coherent fluctuations for a given frequency, in particular the low-frequency shock oscillations and its relation to the recirculating flow region. 
The SPOD is used here as an auxiliary tool to observe the causes of the shock unsteadiness and to visualize the propagation of the fluctuations for a targeted frequency within the SBLI. 
SPOD was employed in a number of studies \citep{picard2000pressure,cavalieri2013wavepackets,semeraro2016stochastic,towne2018spectral,sasaki2020role}, however, to the authors' knowledge, its application to the problem of a shock wave boundary layer interaction has not been reported previously.

The spectral decomposition is applied to the spanwise-averaged streamwise/wall-normal $(u,v)$ velocity fluctuations such that they are optimal modes to represent the turbulent kinetic energy. The modes are defined from the solution of the following integral equation:

\begin{equation}
\label{spodequation1}
\int \mathbf{\Gamma}(\mathbf{x},\mathbf{x'},f)\mathbf{W}(\mathbf{x'})\psi_j(\mathbf{x'},f)\mathrm{d}\mathbf{x'}=\lambda \psi_i(\mathbf{x},f).
\end{equation}

Here, $\mathbf{x}=(x,y)$ and $\mathbf{W}(\mathbf{x'})$ is a weighting matrix which allows one to account more either at arbitrary flow regions or due to the use of non-cartesian grid; $\psi_i$ corresponds to an eigenfunction ($i$th SPOD mode) with corresponding eigenvalue $\lambda$ and $\mathbf{\Gamma}(\mathbf{x},\mathbf{x'},f)$ is the two-point cross spectral density, which is defined from the Fourier transform of the correlation tensor

\begin{equation}
\label{definingthecrosspowerspectraltensor}
\mathbf{\Gamma}(\mathbf{x},\mathbf{x'},f)=\int_{-\infty}^{\infty}\mathbf{C}(\mathbf{x},\mathbf{x'},\tau)e^{i 2\pi f \tau}\mathrm{d}\tau.
\end{equation}

\noindent The correlation tensor $\mathbf{C}$ is obtained by:

\begin{equation}
\label{correlationtensor}
\mathbf{C}(\mathbf{x},\mathbf{x'},\tau)=\langle \mathbf{q}(\mathbf{x},t)\mathbf{q}^*(\mathbf{x}^\prime,t+\tau) \rangle,
\end{equation}

\noindent with $\mathbf{q}=(u,v)$, the velocity components, $*$ representing the conjugate transpose, and $\rangle . \langle$ supplying an ensemble time average, representing the expected value of a given realization of the flow field.

The method of snapshots was used to compute the SPOD modes, where the Welch's method is applied to compute the cross-spectral density.
A detailed description of this procedure can be found in \citet{towne2018spectral} or in the appendix of \citet{sasaki2020role}.
The discretized form of equation \eqref{spodequation1} is replaced by an eigenvalue problem \citep{towne2018spectral}, for each discrete frequency $f_k$, which reads:

\begin{equation}
\label{spodequation2}
\mathbf{\Gamma}_{f_k}\psi_{f_k}=\psi_{f_k}\lambda_{f_k},
\end{equation}

\noindent where $\mathbf{\Gamma}_{f_k}$ refers to the cross-spectral-density matrix and it is calculated via the ensemble averaging

\begin{equation}
\label{spodequation3}
\mathbf{\Gamma}_{ij}=\langle \hat{\mathbf{q}}_i \hat{\mathbf{q}}_j^* \rangle.
\end{equation}

The subscripts $f_k$ are not shown for simplicity, and $\hat{\mathbf{q_i}}=[\mathbf{u}_i, \mathbf{v}_i]^T$, with the superscript $T$ representing a transpose, denotes the $i$-th realisation of the velocity field (i.e., the $i$-th short-time Fourier transform of the velocity field in the Welch's method).

The elements in equation \ref{spodequation3} were determined by means of the Welch's method, with a triangular window and 75\% overlap of the segments. 
Each segment has 5000 points with a time discretization of $\Delta t=0.026$. 
The total number of elements in the time signal considered is of 342900, which results in 97 blocks for averaging. 
These parameters were observed to be adequate to resolve the flow structures in the particularly low frequencies of interest.

\section{Transfer functions in the interaction zone}
\label{predictionresultsforthedata123}

\subsection{Prediction performance}

We start by assessing the accuracy of the single-input single-output linear transfer functions in predicting the velocity/pressure fluctuations in the average shock position. 
The fluctuations were low-pass filtered using a finite-impulse-response (FIR) filter with a cut-off frequency of $St=0.3$ to avoid the influence of higher-frequency flow structures.
The filter is applied prior to the calculation of the transfer functions. 
A window size consisting of 2000 points, which corresponds to $\Delta t=60$ with an overlap of 75\% was used in the Welch's method for calculation of the power spectra.
Similar results were also obtained for window sizes consisting of 5000 points. 
To avoid spurious values of the transfer function, frequencies where $S_{IO}(f)$ is below a certain threshold are set to zero. 
This occurs for frequencies higher than the cut-off filter.

Three input/output pairs were computed: pressure/pressure, velocity/pressure and velocity/velocity. 
Here, only the causal part of the convolution kernel was considered to compute the prediction, \textit{i.e.} $g_{IO}$ was forced to zero when $t<0$. 
To quantitatively evaluate the prediction performance, the normalized correlation between the estimation $O_{est}(t)$ and the LES data $O_{LES}(t)$ was computed

\begin{equation}
\label{equationforcorrelation}
C=\frac{\int_{-\infty}^{\infty}O_{LES}(t)O_{est}(t)\mathrm{d}t}{\sqrt{\int_{-\infty}^{\infty}O_{LES}^2(t)\mathrm{d}t}\sqrt{\int_{-\infty}^{\infty}O_{est}^2(t)\mathrm{d}t}},
\end{equation}

\noindent where $O_{LES}$ and $O_{est}$ are the LES and the estimated data, respectively. 
The normalized correlation in the previous equation varies between zero and one, where one indicates a perfect correlation between predicted and LES signals.

\begin{figure}
\begin{center}
\subfigure[Input upstream of the shock, pressure/pressure estimation]{\includegraphics[width=0.48\textwidth]{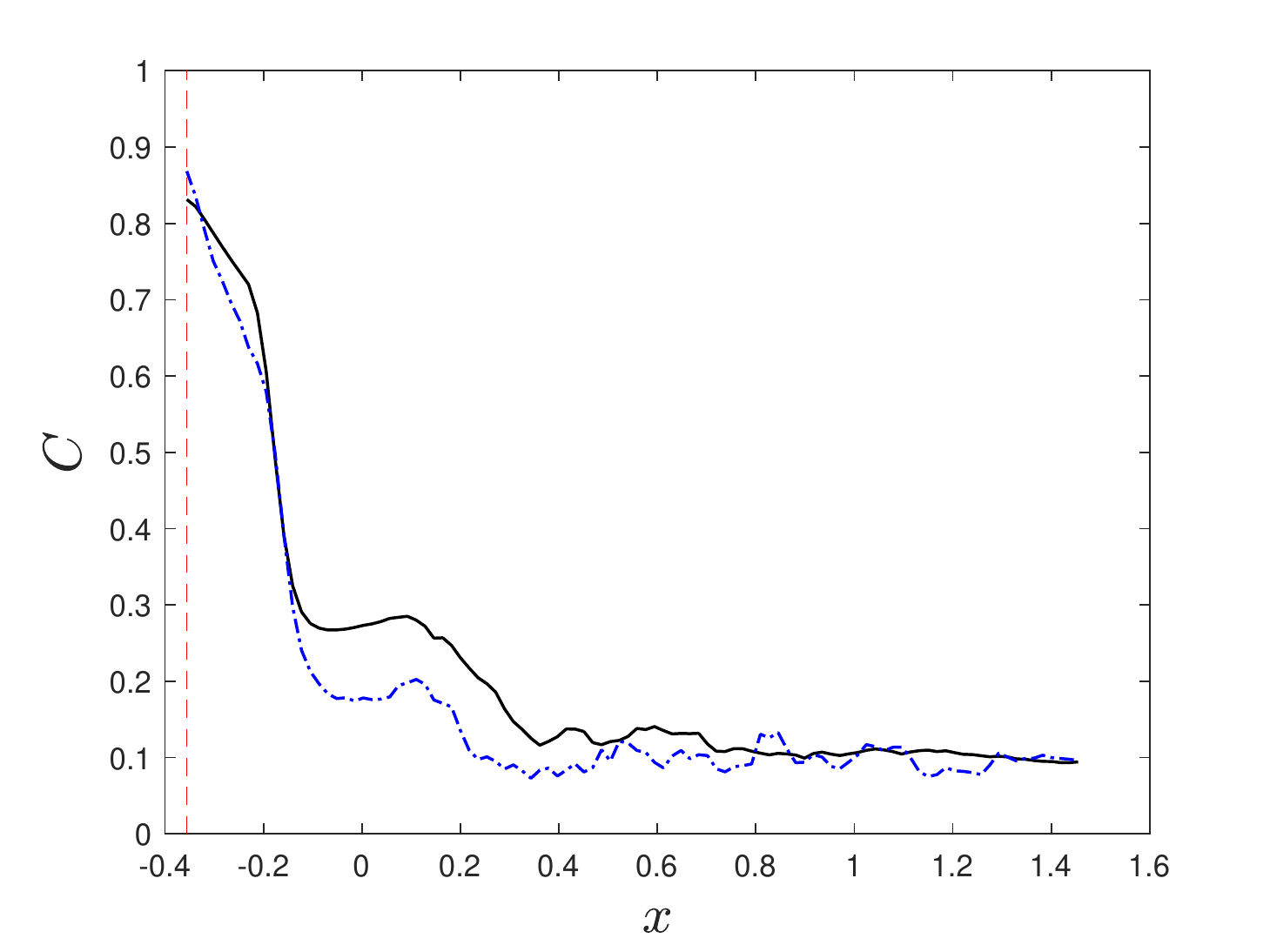}}
\subfigure[Input downstream of the shock, pressure/pressure estimation]{\includegraphics[width=0.48\textwidth]{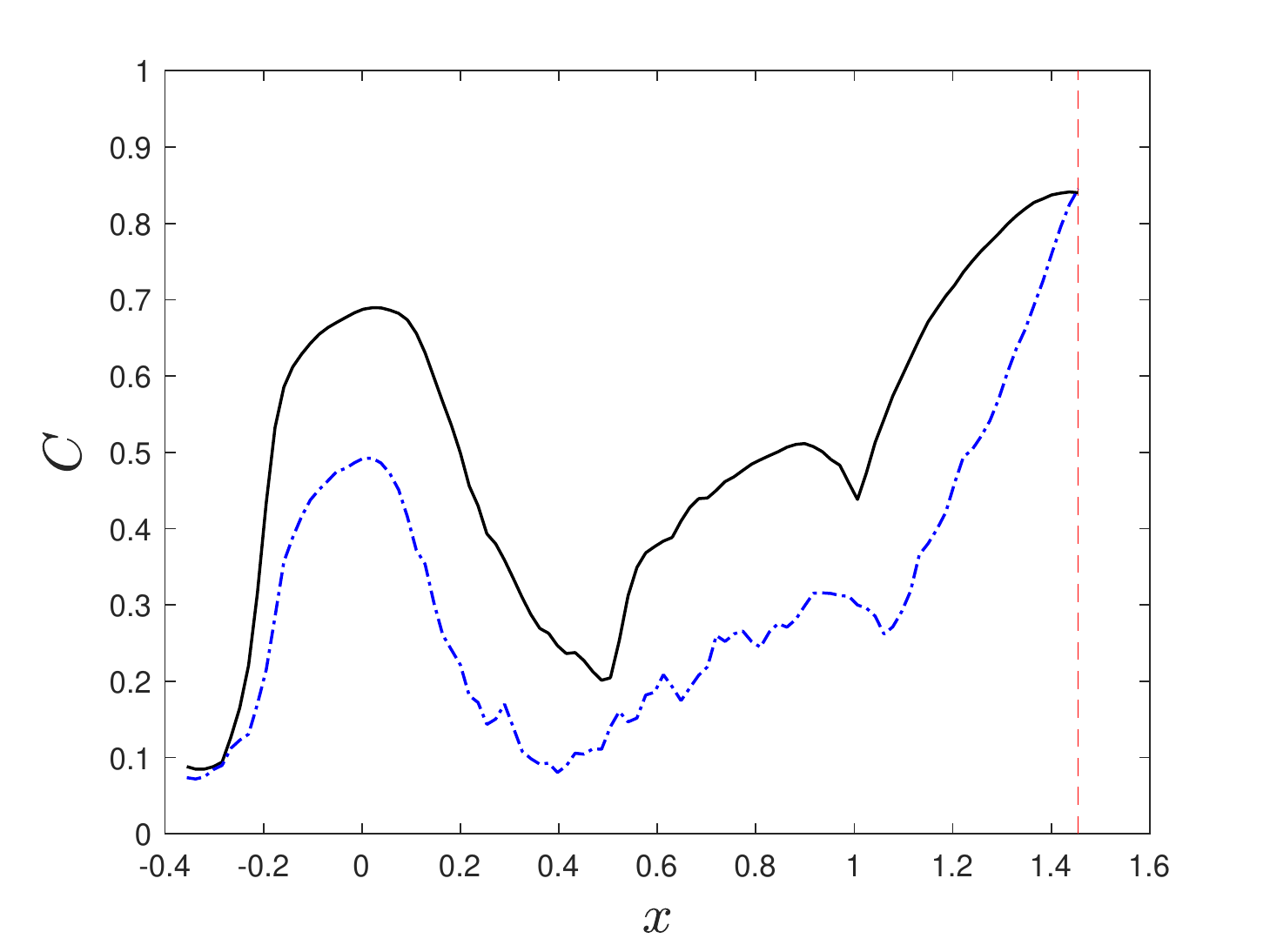}}\\
\subfigure[Input upstream of the shock, velocity/pressure estimation]{\includegraphics[width=0.48\textwidth]{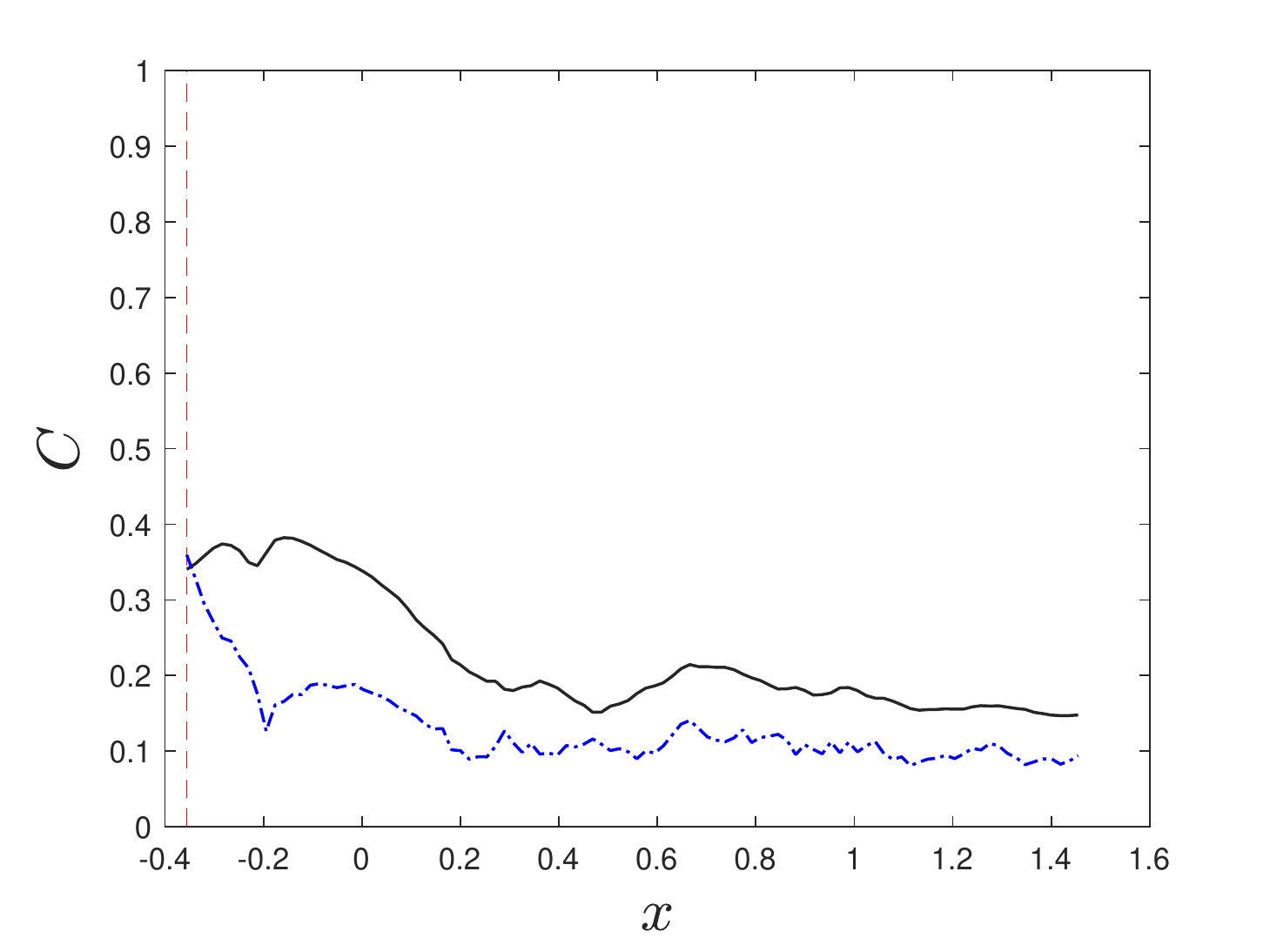}}
\subfigure[Input downstream of the shock, velocity/pressure estimation]{\includegraphics[width=0.48\textwidth]{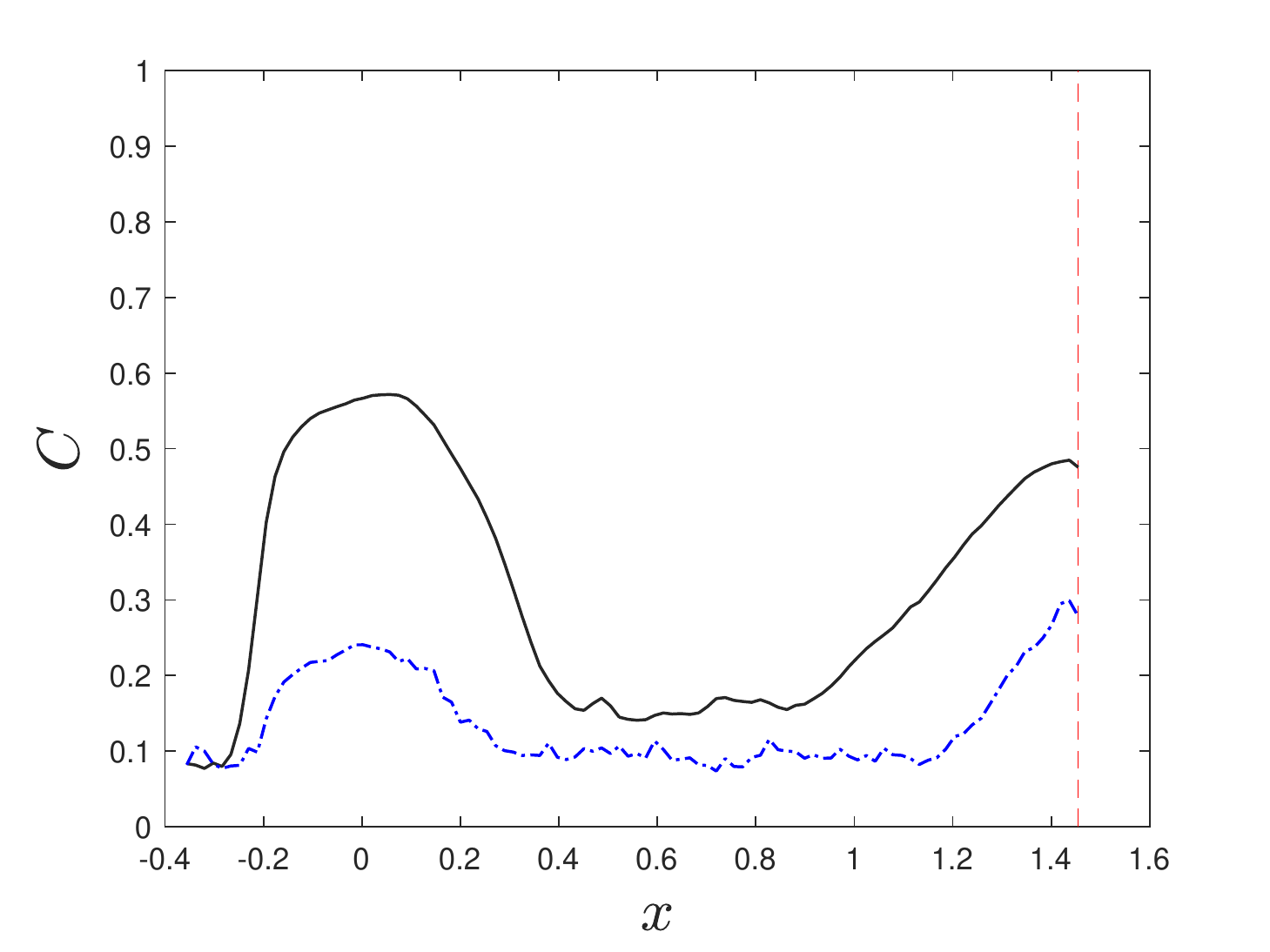}}\\
\subfigure[Input upstream of the shock, velocity/velocity estimation]{\includegraphics[width=0.48\textwidth]{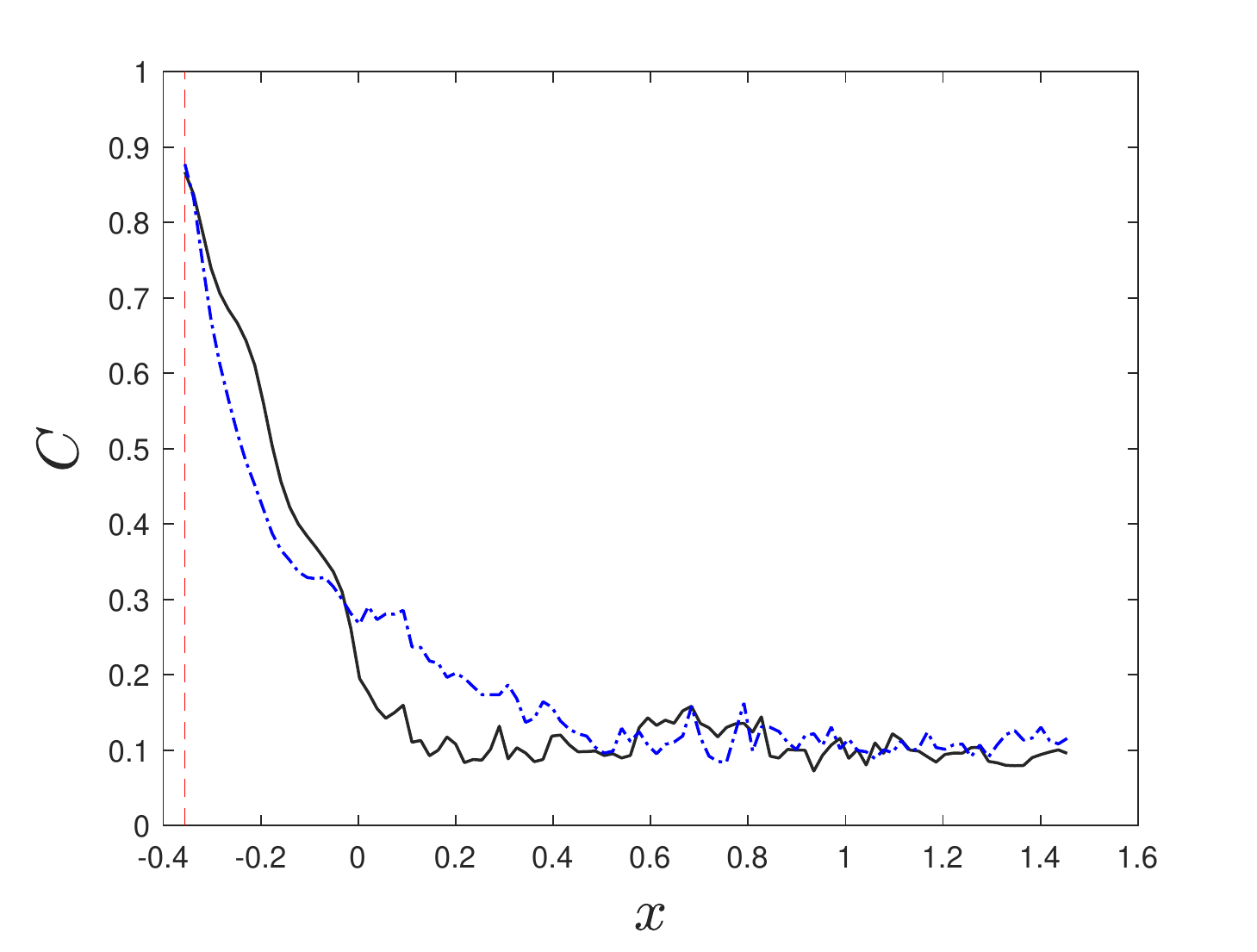}}
\subfigure[Input downstream of the shock,velocity/velocity estimation]{\includegraphics[width=0.48\textwidth]{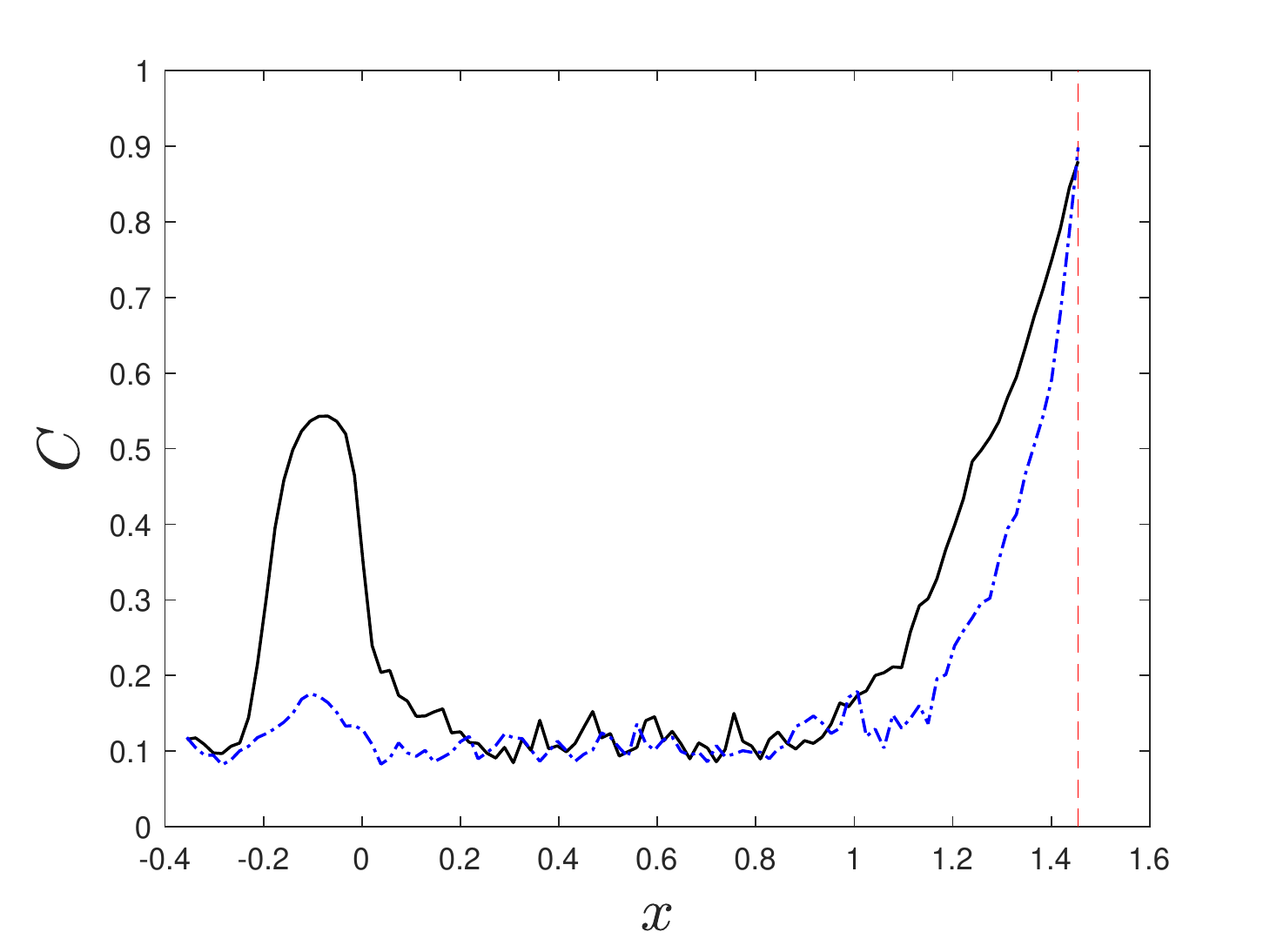}}\\
\end{center}
   \caption{Correlation between prediction and LES for pressure and streamwise velocity components. The vertical dashed line presents the position of the input. Solid black lines correspond to the prediction performed with the spanwise averaging and dash-dot blue lines without it.}
  \label{examplecorrelationsfasdf}
\end{figure}

Two input positions were considered separately, upstream and downstream of the shock.
The correlation was computed as a function of the output (prediction) streamwise position. 
Figure \ref{examplecorrelationsfasdf} shows the resulting correlation between prediction and LES data, the vertical line highlights the input location. 
The prediction was performed with and without the spanwise averaging of the input/output quantities.

As expected, the correlations decay as the distance between input and output positions increases; for the cases where the same variable is used as the measurement and prediction, the correlations depart from large values close to unity. 
When the input is upstream of the shock in figures 4(a,c,e), the correlation decays monotonically up to the mean shock location $x=0$ and stabilizes further downstream.
On the other hand, in the case of an input downstream of the shock wave shown in figures 4(b,d,f), the correlation also decays as the output position is moved further away from the input. 
However, close to the shock location, there is a swift increase in the correlation values. 
It should also be observed that even though there is a large separation between input and output, the correlation between prediction and input reaches values above 0.6.

The effect of the spanwise averaging is remarkable, leading to a significant increase in the correlation of the predicted signals, in particular when the input position is taken downstream of the shock motion. 
This is in accordance with the spectra of figure \ref{2DSpectra}, which indicates a predominantly two-dimensional nature of the streamwise velocity shock-related fluctuations. 
Hereafter, only spanwise averaged data is considered.

\begin{figure}
\begin{center}
\subfigure[Input downstream of the shock, pressure/pressure estimation]{\includegraphics[width=0.75\textwidth]{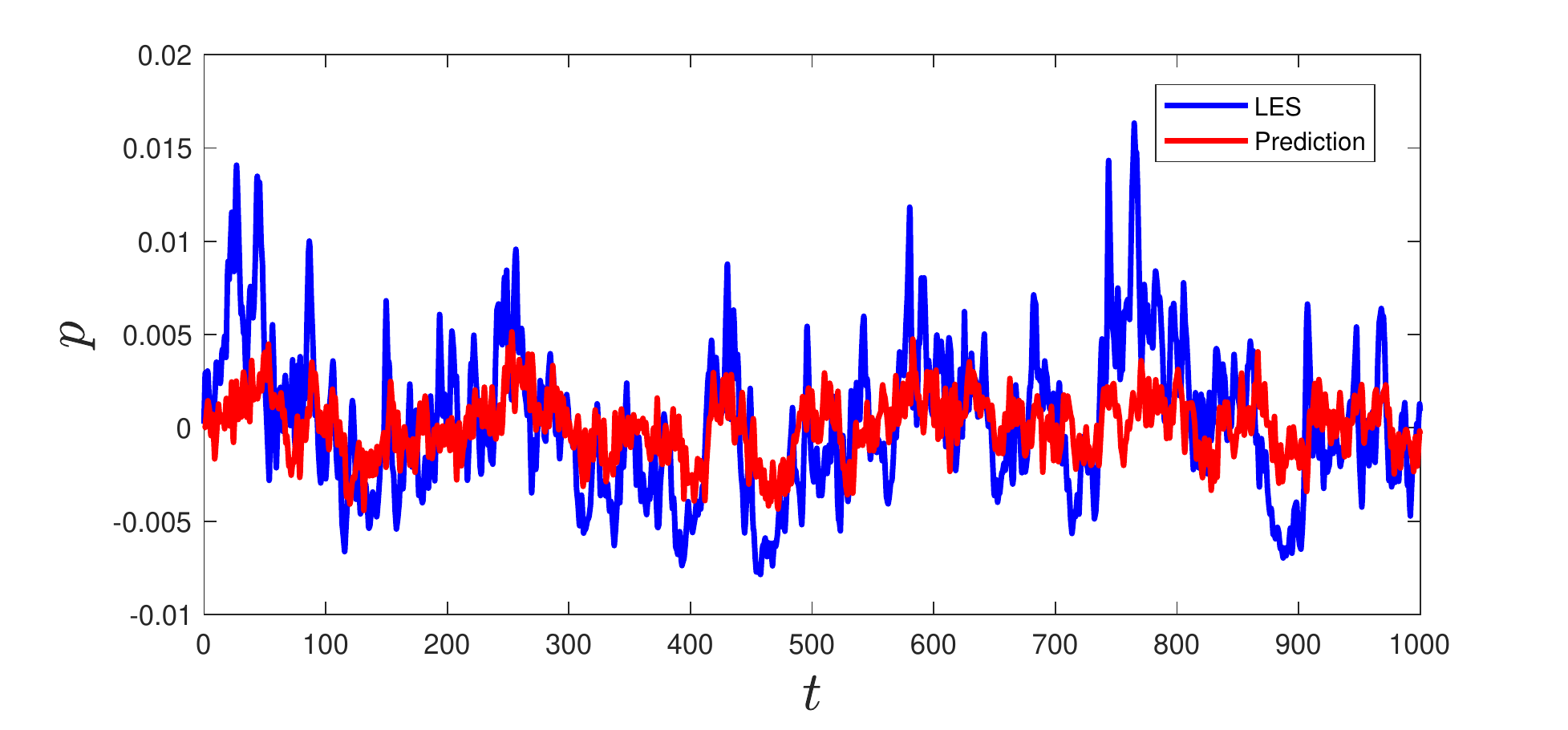}}\\
\subfigure[Input downstream of the shock, velocity/pressure estimation]{\includegraphics[width=0.75\textwidth]{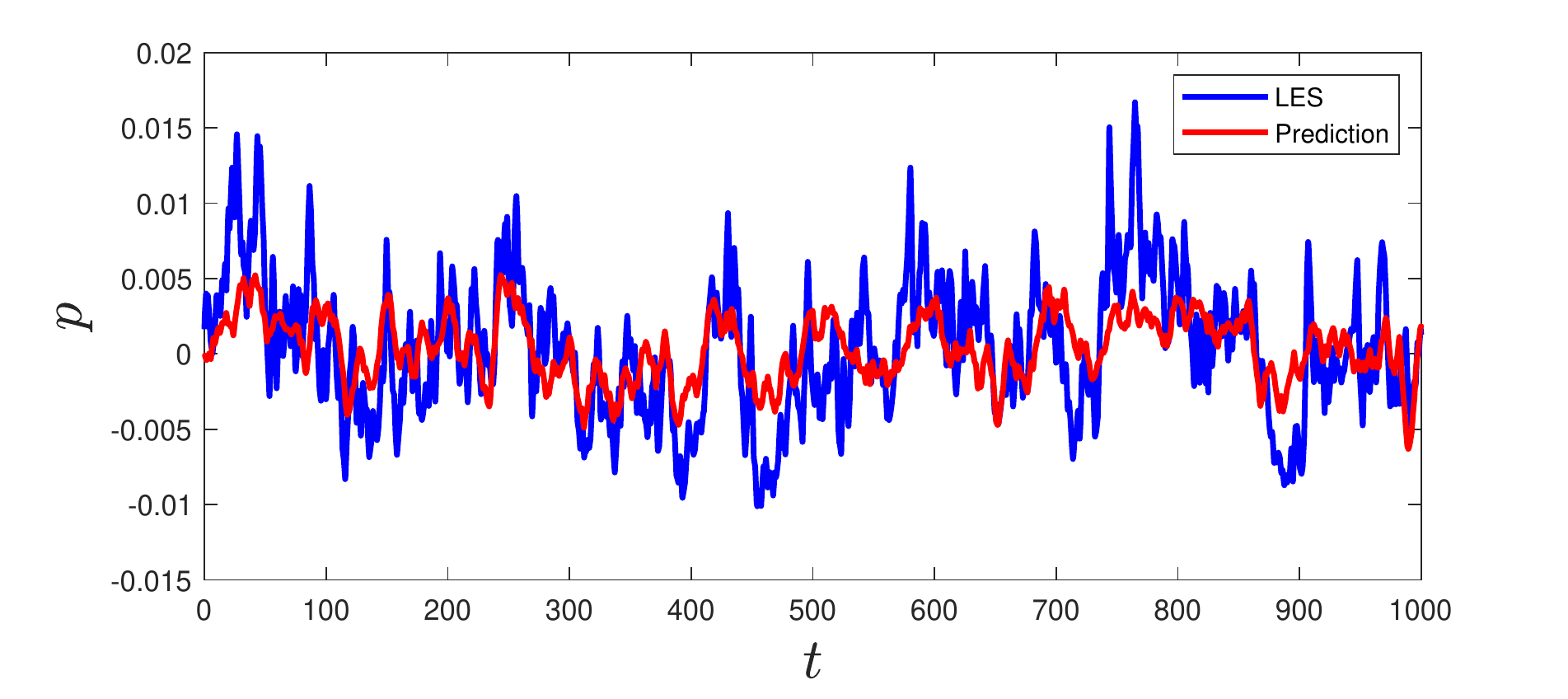}}\\
\subfigure[Input downstream of the shock, velocity/velocity estimation]{\includegraphics[width=0.75\textwidth]{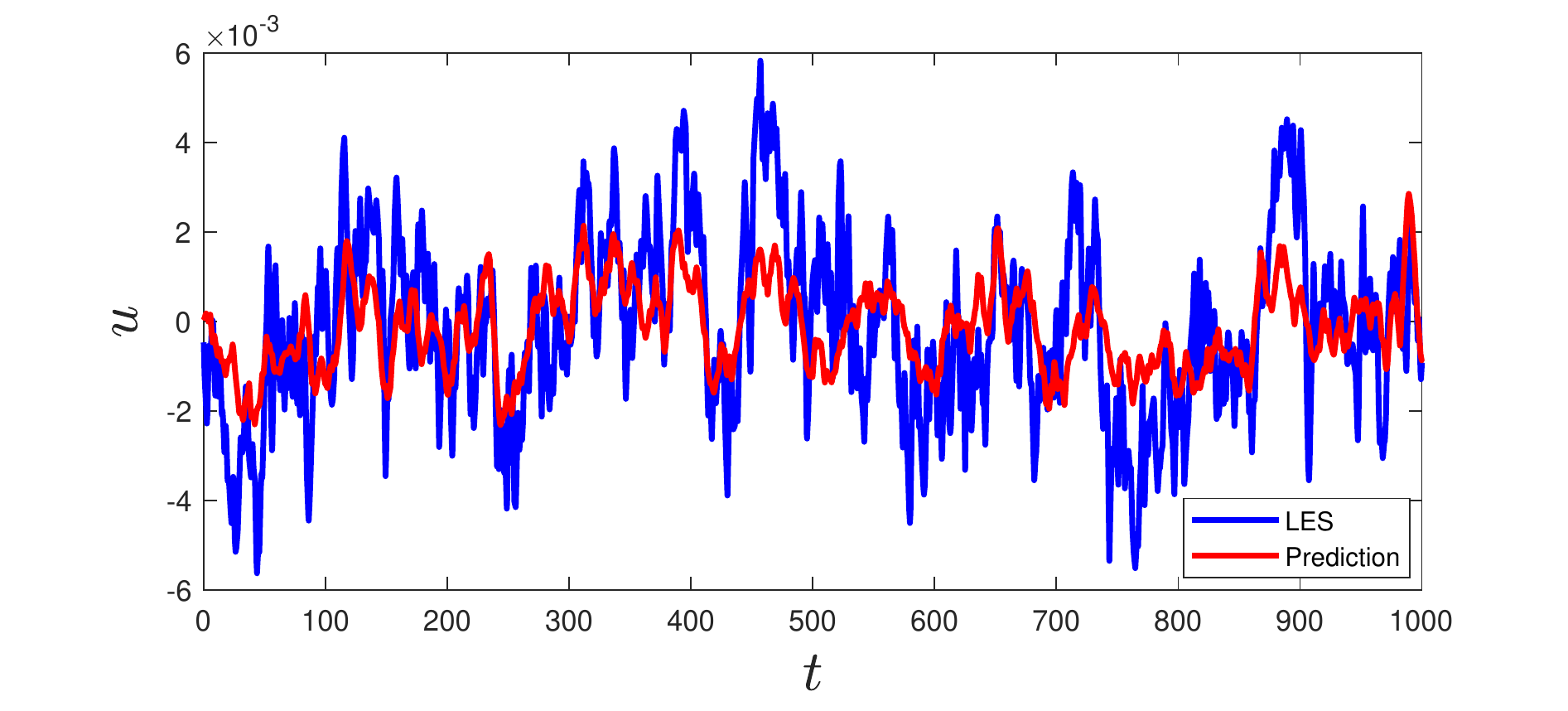}}\\
\end{center}
   \caption{Comparison between the time signals from LES and linear prediction, for pressure (a) and streamwise velocity (b) fluctuations. The predicted position corresponds to $x=0$ (average position of the shock) and the input is located at $x=1.5$ downstream of the shock.}
  \label{examplepredictionstimedomain}
\end{figure}

Figure \ref{examplepredictionstimedomain} presents a sample of the predicted streamwise velocity and pressure fluctuations at $x=0$ with input at $x_{in}=1.5$. 
It is noticeable that, even though the amplitude of the LES signal is underestimated, both the phase and the low-frequency content of the signal (characteristic of the shock motion) are well captured by this approach, in spite of the large streamwise separation between input and output. The corresponding shape of the transfer function is depicted in figure \ref{exampleoftfdata}.

\subsection{Wall-normal behaviour and causality of the transfer function predictions}

To further investigate the prediction performance and to track the causal areas for the shock fluctuations, the output position was fixed at $x=0$, near the wall. 
Transfer functions were computed considering a grid of positions along the streamwise and wall-normal directions as input signals. 
As indicated above, we consider the spanwise averaged velocity fields low-pass filtered with the cut-off frequency $St=0.3$ prior to the computation of the transfer functions. 
The same window size and overlap of the previous analysis were used for calculating the power spectra. 
The pressure fluctuation was considered as the estimated variable, with the streamwise velocity fluctuations as the input.
Finally, the full convolution kernel including its non-causal counterpart was used here to evaluate signal reconstruction.

Figure \ref{pparameterandgroupvelocity}(a) presents the resulting correlations between the shock prediction and LES data for the different inputs considered. 
High correlations appear in the near-shock region. 
This is expected due to the coherent shock motion; such high correlations are due to an input related to shock oscillations predicting the fluctuations at the shock foot. 
An interesting behavior occurs downstream of the shock ($x>1.0$), where correlations of the order of 0.8 start to occur, indicating that these can be used to accurately estimate the shock motion. 
In the region $0.2<x<1.0$, within the recirculation bubble, the correlations drop to low values, a feature which can be assessed by the shape of the SPOD modes presented in following section.

\begin{figure}
\centering
\includegraphics[width=0.75\textwidth]{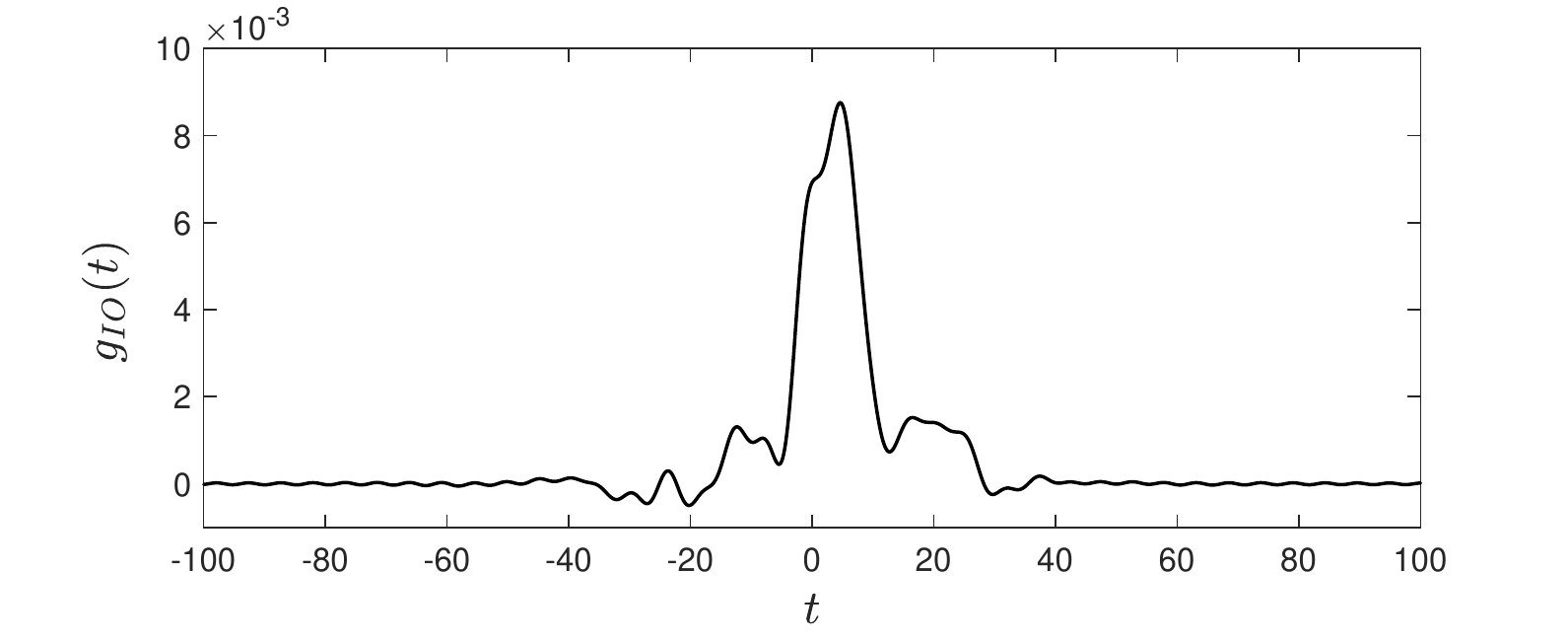}
\caption{Transfer function computed streamwise velocity fluctuation at the positions of $x_{in}=1.5$ and $x_{out}=0$.}
\label{exampleoftfdata}
\end{figure}

\begin{figure}
\begin{center}
\subfigure[Correlations between estimated and LES data]{\includegraphics[width=0.80\textwidth]{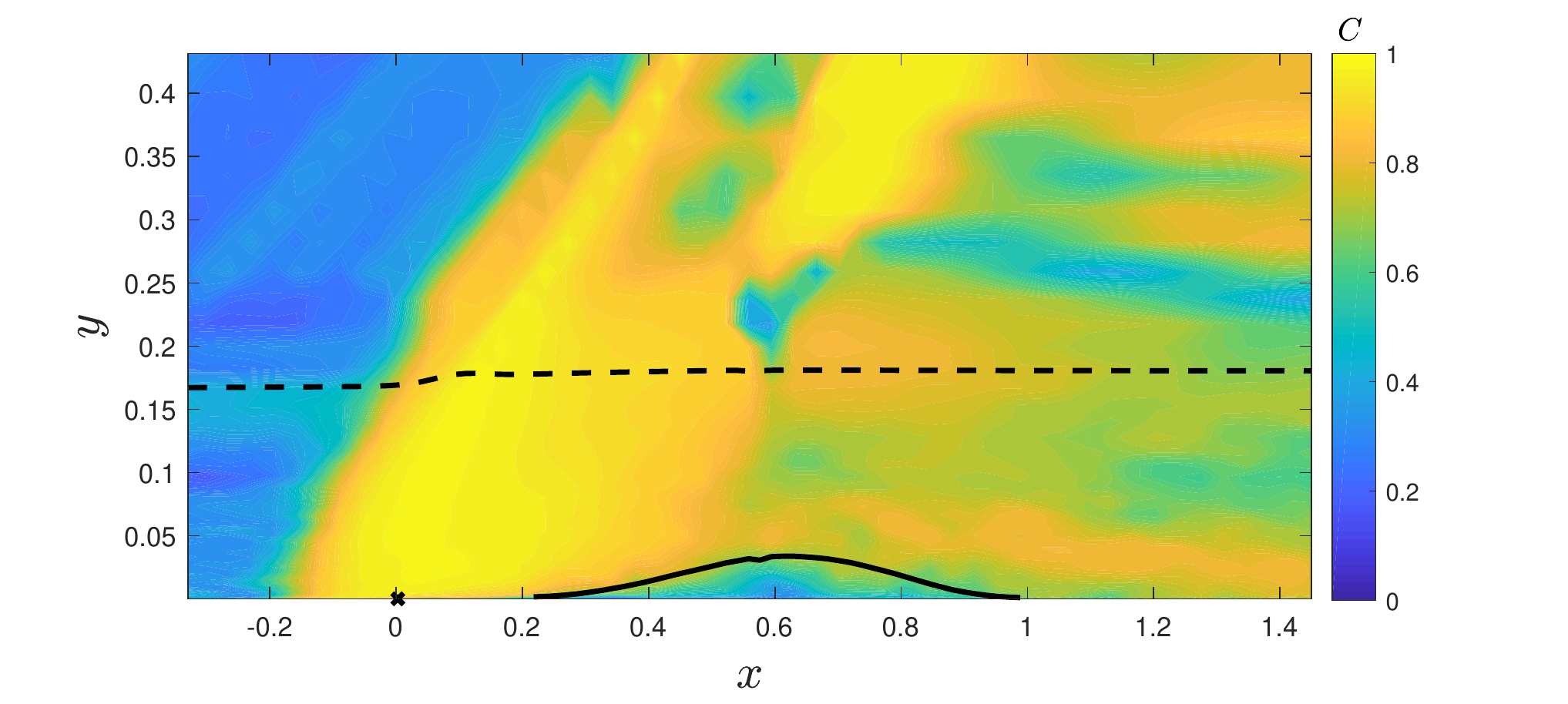}} \\
\subfigure[$P-$paramter]{\includegraphics[width=0.80\textwidth]{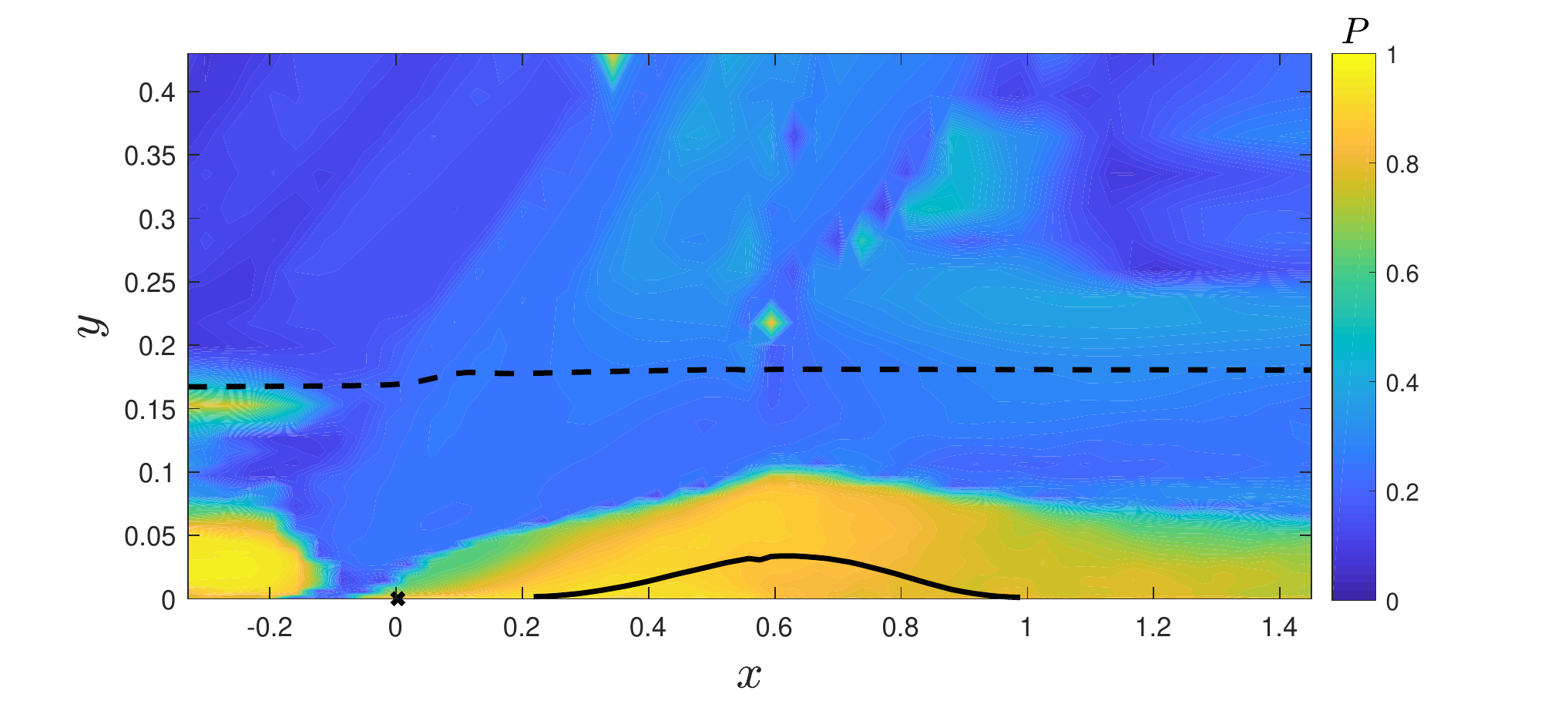}}\\
\subfigure[Estimated group velocity]{\includegraphics[width=0.80\textwidth]{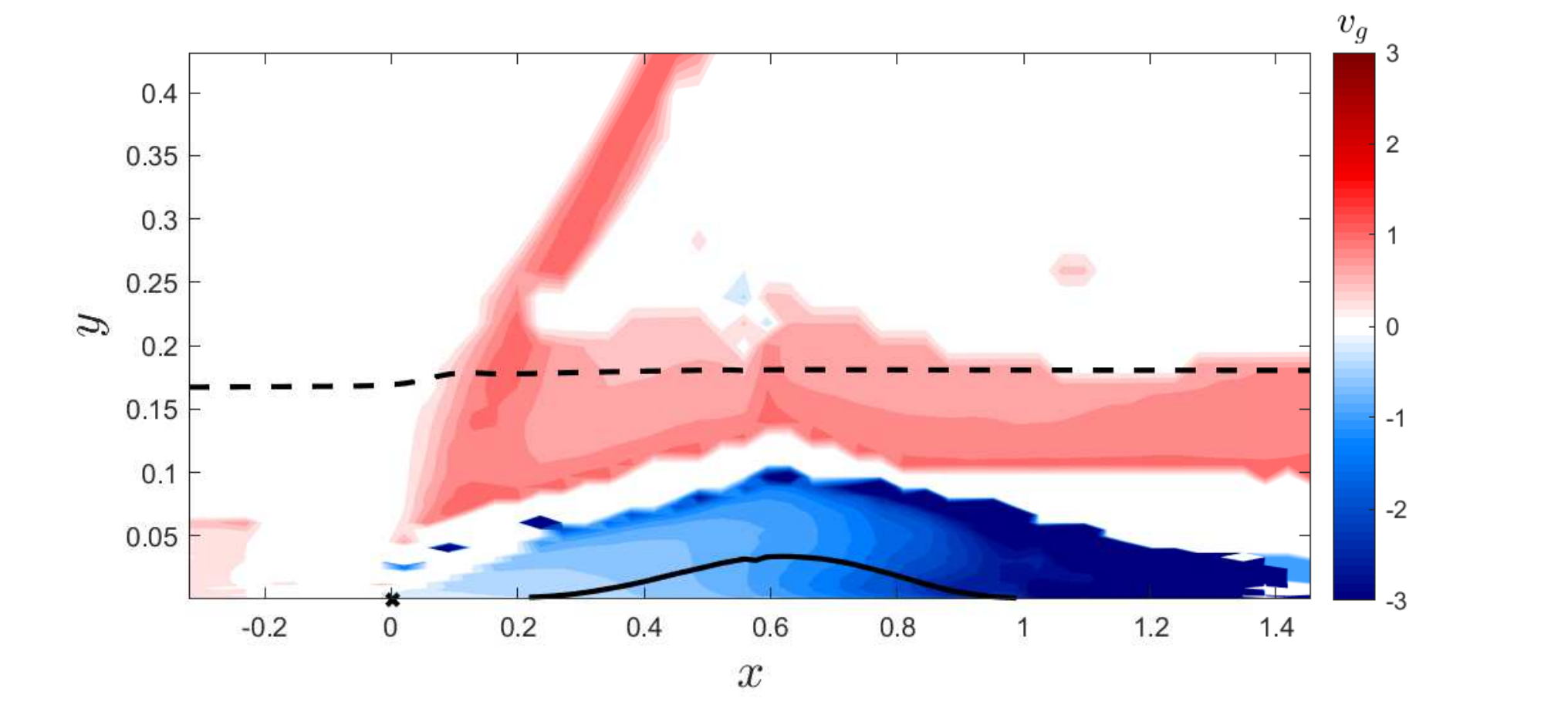}}
\end{center}
   \caption{Effect of the input position for prediction. Correlation values between predicted signal and LES data for the pressure fluctuations (a), $P-$parameter for computing the causal relationship (b) and streamwise velocity and estimated group velocity (c). The solid line indicates the recirculation bubble and the dashed line the boundary layer thickness. The cross indicates the considered output position}
  \label{pparameterandgroupvelocity}
\end{figure}

In order to determine the causal effects of the shock, figures \ref{pparameterandgroupvelocity}(b,c) present two indices, the $P-$parameter, related to the causality between input and output positions, and the group velocity. 
The latter can be estimated from the time delay observed in the empirically calculated transfer function and the streamwise separation between input and output positions. 
As an example, consider the time-domain transfer function computed using the streamwise velocity fluctuation at the positions of $x_{in}=1.5$ and $x_{out}=0$, shown in figure \ref{exampleoftfdata}. 
The streamwise separation is of -1.5, with a time delay for the peak of the transfer function corresponding to approximately $\Delta t=2$, which results in an estimated group velocity of $v_g=-0.75$, normalized with respect to the free-stream velocity $U_\infty$. 
As the bulk of the transfer function is for positive time delays, the causality parameter $P$ is calculated as 0.9, indicating a nearly causal behavior. 
Regions where the the $P$-parameter was between 0.3 and 0.7, or where the correlation $C$ between estimation and prediction was below 0.3, were not considered in the computation of the group velocity and the value was set to $v_g=0$.
This avoids a non-physical interpretation of this parameter near the $v_g=0$ line or where the transfer function estimation is inaccurate.

When both the correlation $C$ and $P-$parameter in figures \ref{pparameterandgroupvelocity}(a,b) are analyzed together, we note that the only causal input locations (large $P$) that lead to high correlations with shock motion (large $C$) surrounds the recirculation bubble and extends downstream of it. 
In particular, the parameter $P$ and the group velocity indicate causality between shock oscillations at $x=0$ from fluctuations near the reattachment point and downstream of the recirculation bubble. 
The computed negative group velocity points to fluctuations that are propagating upstream from the input location.
In addition, the high correlation values indicate that fluctuations at downstream locations precede shock oscillations, and enable an accurate prediction of its motion. 

Fluctuations occurring upstream of the shock present a positive group velocity and are not strongly correlated with its motion. 
These results are in agreement with previous observations from \citet{Touber2009}, who observed a negative phase speed in a similar configuration, and related it to the existence of a global mode. 
The observations from \citet{Dupont2006} and \citet{Wu2008} also point to a strong correlation between low-frequency fluctuations and the reattachment region.
It is of interest to remark the existence of a region upstream of the shock with positive group velocity and large values of $P$. 
This region contributes to causing the shock motion, which is in accordance with the experiments from \citet{Ganapathisubramani2007}. However, due to the relatively low correlation levels presented in figure \ref{examplecorrelationsfasdf}(a), this region has a relatively smaller contribution to the shock unsteadiness than the downstream zone in the present flow simulation.

\section{Upstream travelling waves detected via SPOD}

\subsection{Shape and velocity of the spectral modes}

\begin{figure}
\centering
\includegraphics[scale=0.65]{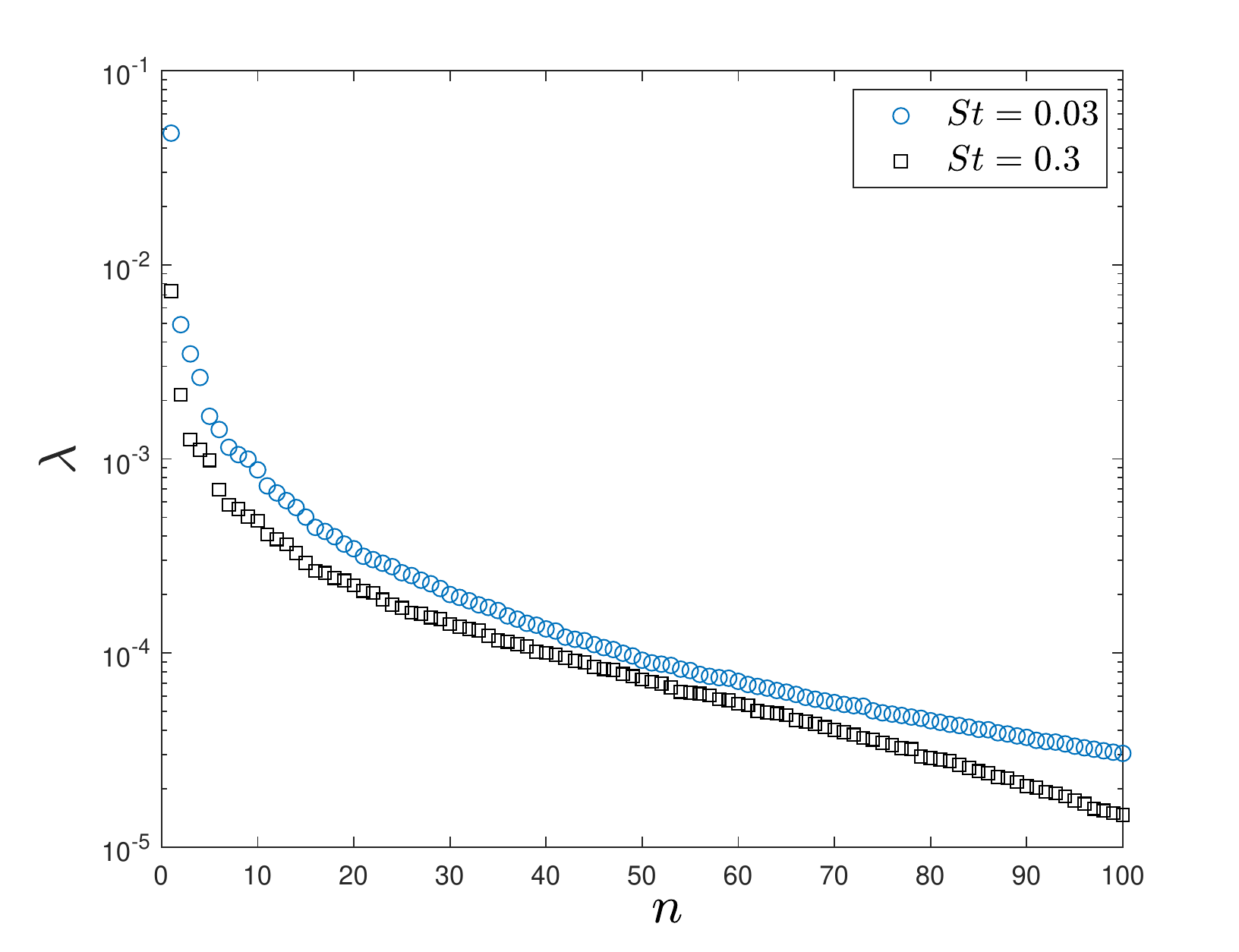}
\caption{Resulting SPOD eigenvalues computed for $St=0.03$ (blue circles) and $St=0.3$ (black squares).}
\label{evaluationsofspodeigenvalues}
\end{figure}

In order to better understand the spatial structure of the low-frequency oscillations, SPOD modes of the spanwise averaged $(u,v)$ velocity components were computed at $St=0.03$ and $St=0.3$. 
The first Strouhal number corresponds to the low-frequency motion, and is representative of shock oscillations, whereas the second one is related to high-frequency boundary-layer and shear-layer fluctuations mostly uncorrelated to the shock. 
Figure \ref{evaluationsofspodeigenvalues} depicts the resulting eigenvalues for these two selected frequencies. 
In both cases, there is a clear dominance of the first mode with respect to the other spectral modes. 
It is also noticeable that the low-frequency eigenvalues are higher over all the spectra. 
Hence, we consider only the first mode in the following analysis. 

\begin{figure}
\begin{center}
\includegraphics[scale=0.6]{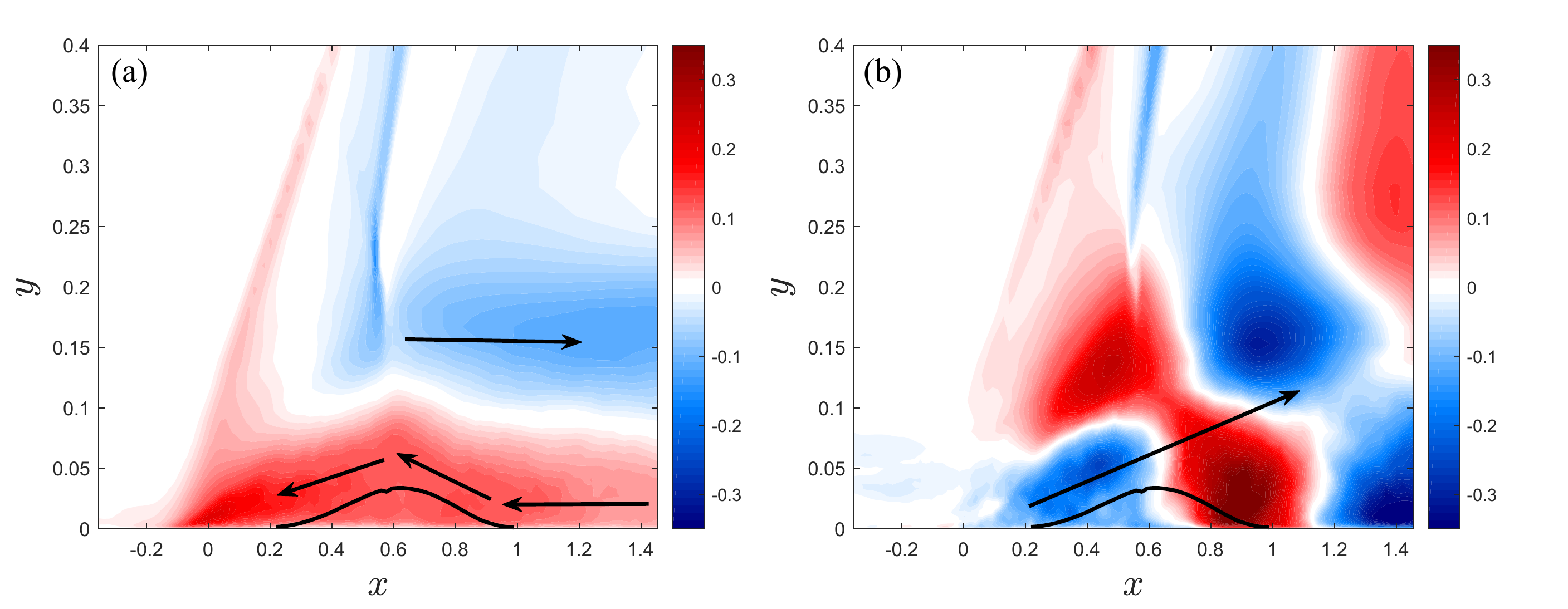}
\end{center}
   \caption{Snapshot of the first SPOD mode of the streamwise velocity component at $St=0.03$ (a) and $St=0.3$ (b). The arrows depict schematically the motion of the streamwise velocity fluctuations. The contours correspond to the real part of the first SPOD mode. The time evolution of both modes is reported in the supplemental material. }
  \label{SPOD_modes}
\end{figure}

Figure \ref{SPOD_modes} presents the structure of the streamwise velocity fluctuations of the first SPOD mode for the frequencies $St=0.03$ and $St=0.3$. 
The time evolution of the corresponding modes were obtained by multiplying the spectral mode by $e^{i 2 \pi f t}$.
The resulting snapshots are reported in the the supplemental movies. 
Interestingly, a wave propagating with negative phase velocity (from right to left) is observed for $St=0.03$, on a position near the wall. Such motion surrounds the recirculation bubble toward the shock foot. 
At a higher wall-normal position ($y \approx 0.15$) there is a wave generated on the shock, propagating with positive group velocity (from left to right). 
The spatial footprint of this upstream travelling motion corresponds well to the causal locations presented in figure\ref{pparameterandgroupvelocity} (b). 
This motion is schematized with the arrows in figure \ref{SPOD_modes}(a), and occurs for low-frequency modes in the vicinity of $St=0.03$. 
For $St=0.3$, only a wave travelling with positive phase velocity appears. 

\begin{figure}
\begin{center}
\subfigure[]{\includegraphics[width=0.48\textwidth]{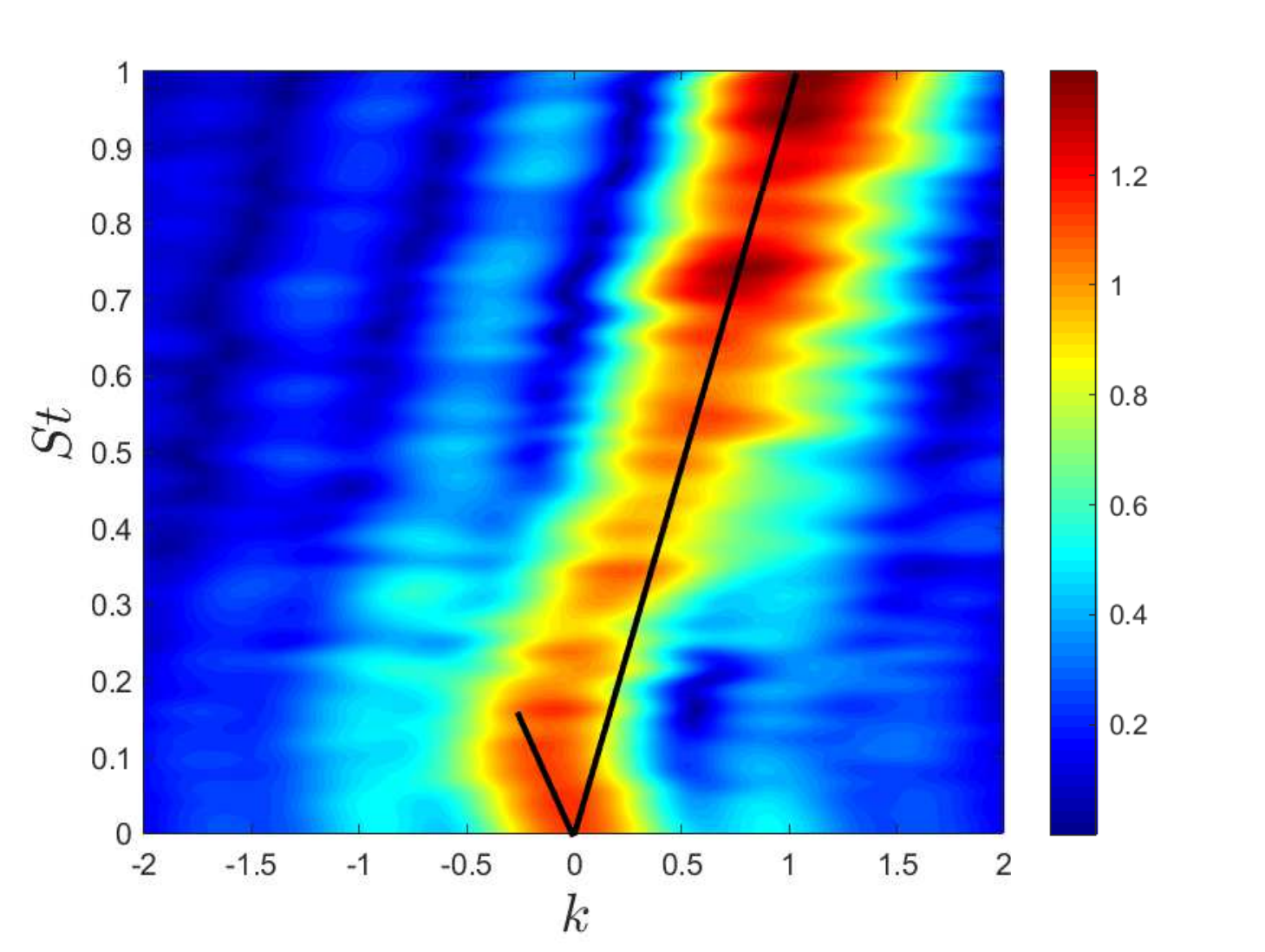}}
\subfigure[]{\includegraphics[width=0.48\textwidth]{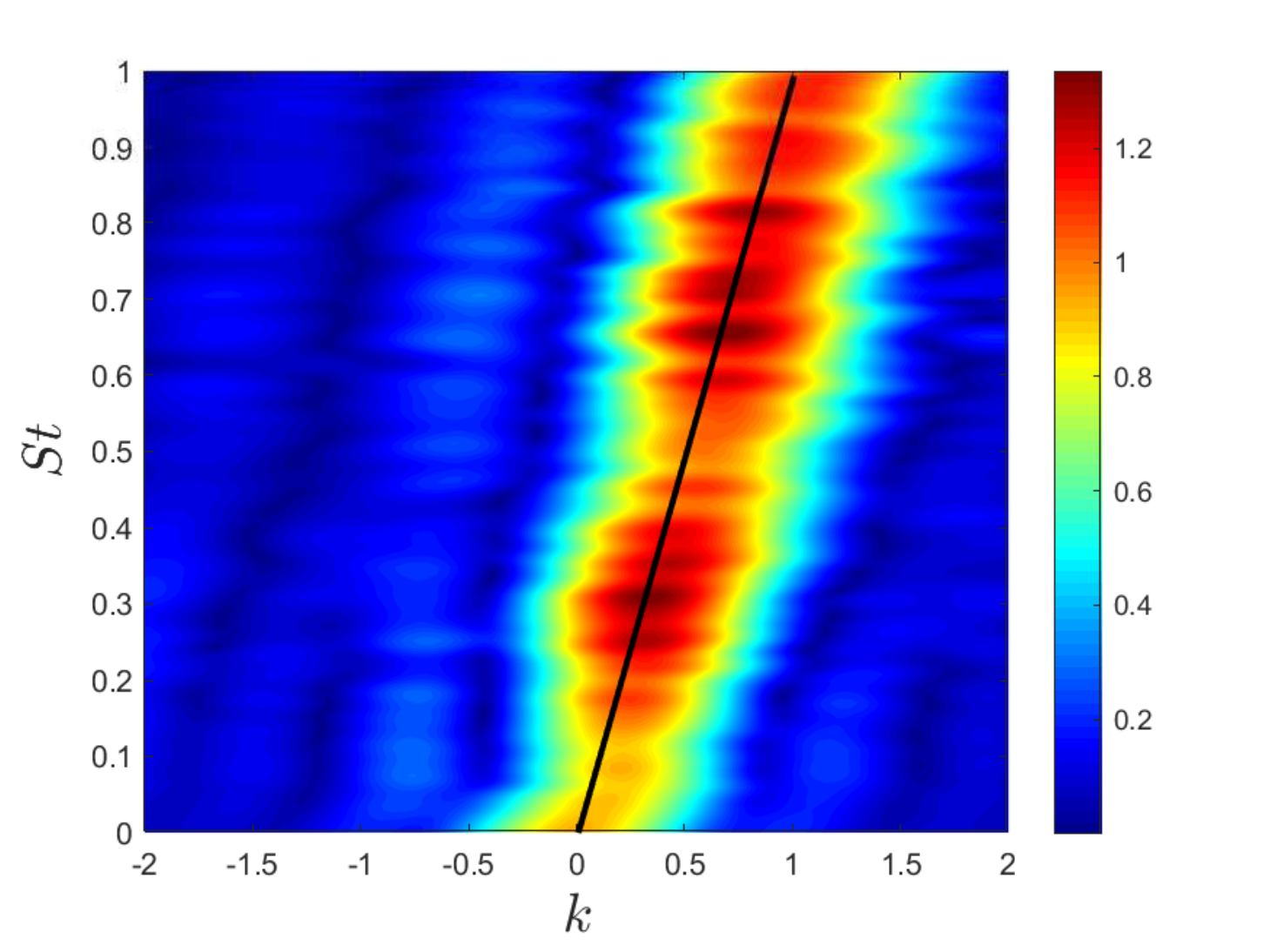}} \\
\end{center}
   \caption{Empirical dispersion relations, computed from the first SPOD mode of the streamwise velocity at $y=0.05$ (a) and $y=0.15$ (b). The logarithmic of the resulting power spectral density was taken. The solid vertical lines indicate constant group velocities for the fluctuations.}
  \label{empiricaldispersionrelationusingu}
\end{figure}

To shed further light on the dynamics of the upstream and downstream travelling waves, an empirical dispersion relation is calculated. 
This was obtained by computing the Fourier transform of the first SPOD mode ($u$) in the streamwise direction as a function of frequency $St$.
A similar approach was employed to study acoustic resonances in turbulent jets \citep{towne_cavalieri_jordan_colonius_schmidt_jaunet_bres_2017}. 
The dispersion relation was computed for two wall-normal positions: $y=0.05$, close to the wall, but surrounding the recirculation bubble, and at $y=0.15$. 
This procedure leads to the frequency-wavenumber power-spectral density diagram of figure \ref{empiricaldispersionrelationusingu}. 
Finally, in order to better resolve the fluctuations occurring at different scales, the iso-contours are plotted in logarithmic scale.

The phase velocity of the fluctuations at a given $(St,k)$ pair is given by $v_p= 2\pi f/k$. 
Therefore, waves with negative values of $k$ propagate upstream. 
Since the group velocity reads $v_g=2\pi\partial f/\partial k$, we note that $v_g$ changes between the two considered wall-normal locations. For the low frequencies ($St<0.1$), there is a group of waves propagating upstream, with negative group velocity and this only occurs in the near-wall region. 
On the other hand, at higher frequencies, the disturbances propagate with positive group velocity independently of the wall-normal position. The waves that travel with negative group velocity contribute to the causality relation between downstream areas and the shock motion, which was quantified by the correlations and the $P$-parameter obtained from the transfer functions. 

\subsection{Reconstruction of the shock motion via SPOD}

In this section, we evaluate how much of the shock oscillations can be recovered from a limited set of data, specifically selected from regions upstream or downstream of the approximate shock position. 
To quantify this analysis, we define the ratio

\begin{equation}
\label{equationforspodreconstruct}
\frac{|u_{SPOD_{cut}}|^2}{|u_{SPOD_{full}}|^2}=\frac{\int_{\Omega}\sum_{i=1}^{N_b}(\lambda_{cut_i}\psi_{cut_i})^2\mathrm{d}x}{\int_{\Omega}\sum_{i=1}^{N_b}(\lambda_{full_i}\psi_{full_i})^2\mathrm{d}x},
\end{equation}

\noindent where $\Omega$ is a horizontal line close to the wall encompassing the shock foot location, and the subscripts $full$ and $cut$ represent SPOD modes which were computed from the whole field and from a limited part of it using the weighting matrix $W$, respectively. Equation \ref{equationforspodreconstruct} represents the ratio between the energy of the streamwise velocity fluctuations from SPOD modes computed using partial knowledge of the flow field and all the data available.

When the whole field is used in the computation, the weighting matrix $W$ considers only the numerical quadrature corresponding to the use of a non-cartesian coordinate system. 
To select specific flow regions, the numerical quadrature is multiplied by the matrix $W_1$, where $W_1=1$ for the selected region in the SPOD calculation, and $W_1=0$ in the region excluded from the calculation. 
Therefore, the SPOD modes are computed considering only a portion of the field. 
In spite of this fact, the whole field can be reconstructed; areas which were not accounted in $W_1$ but are observed in the SPOD modes can be related to fluctuations within the selected region where $W_1=1$. 
This occurs due to the spatial coherence for the frequency where the SPOD mode is calculated. 
We attempt to evaluate through SPOD in regions that exclude the shock itself but are coherent with the shock motion.
A similar approach was performed in \citet{sano2019trailing} to reconstruct the acoustic field scattered by an airfoil from knowledge of its near field fluctuations.

\begin{figure}
\centering
\includegraphics[width=0.80\textwidth]{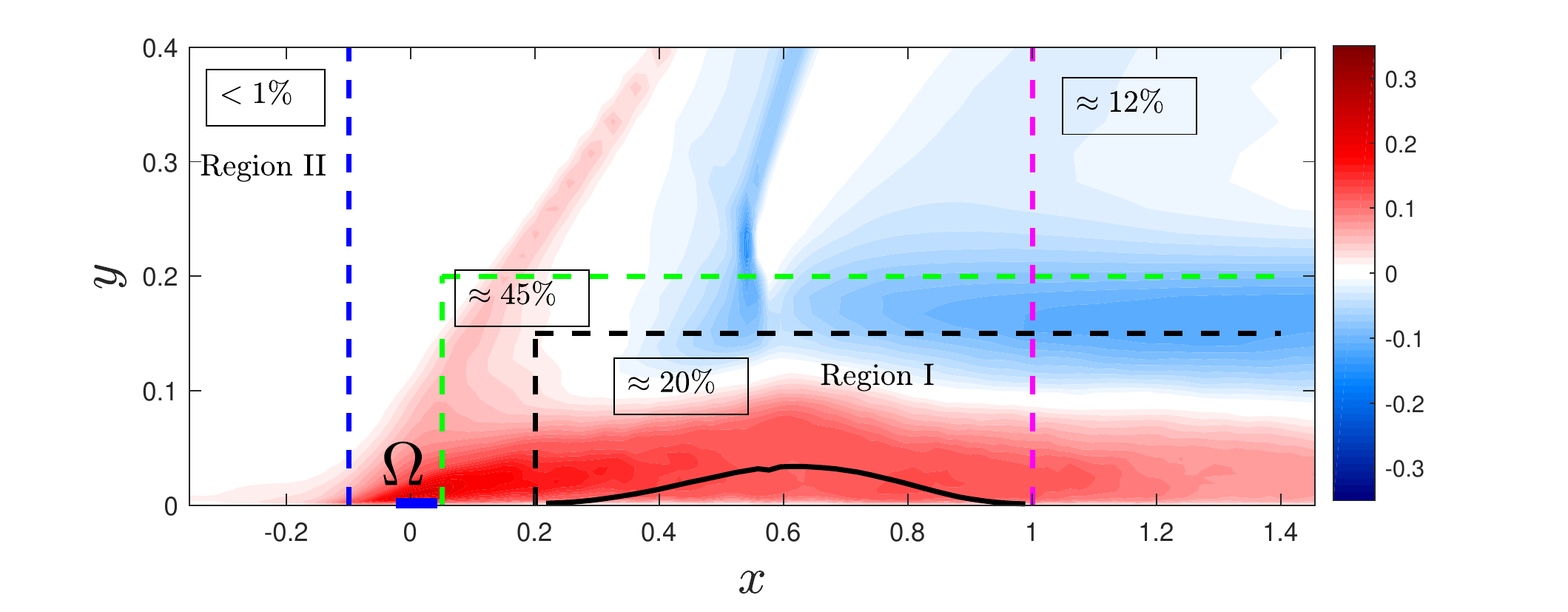}
\caption{First SPOD mode for the streamwise velocity at $St=0.03$ and selected flow regions for the specific SPOD calculations. The portion of the energy at $\Omega$ which can be recovered in each case, in comparison to that of the full field, is highlighted in the corresponding rectangles.}
\label{exampleofareasinspodcalculation}
\end{figure}

Figure \ref{exampleofareasinspodcalculation} provides distinct areas under evaluation. 
The energy is integrated over $\Omega$, a straight line in the first grid position above the wall along $-0.05<x<0.05$.
This line is shown in the plot.
The spectral modes were calculated considering fluctuations within each of the rectangles limited by the dashed lines. 
For each region, $W_1=1$ inside the rectangles, and zero outside them. 
Two rectangles are highlighted, indicating regions I and II, a rectangle downstream (black) and upstream (blue) of the shock, respectively. In the following, these two regions are considered in detail.
The amount of energy that can be recovered is also shown in the figure \ref{exampleofareasinspodcalculation}. 
It is clear that most of the energy of the shock fluctuations is related to downstream fluctuations. 
When considering the upstream region only, less than 1\% of the energy is recovered.

\begin{figure}
\begin{center}
\includegraphics[scale=0.7]{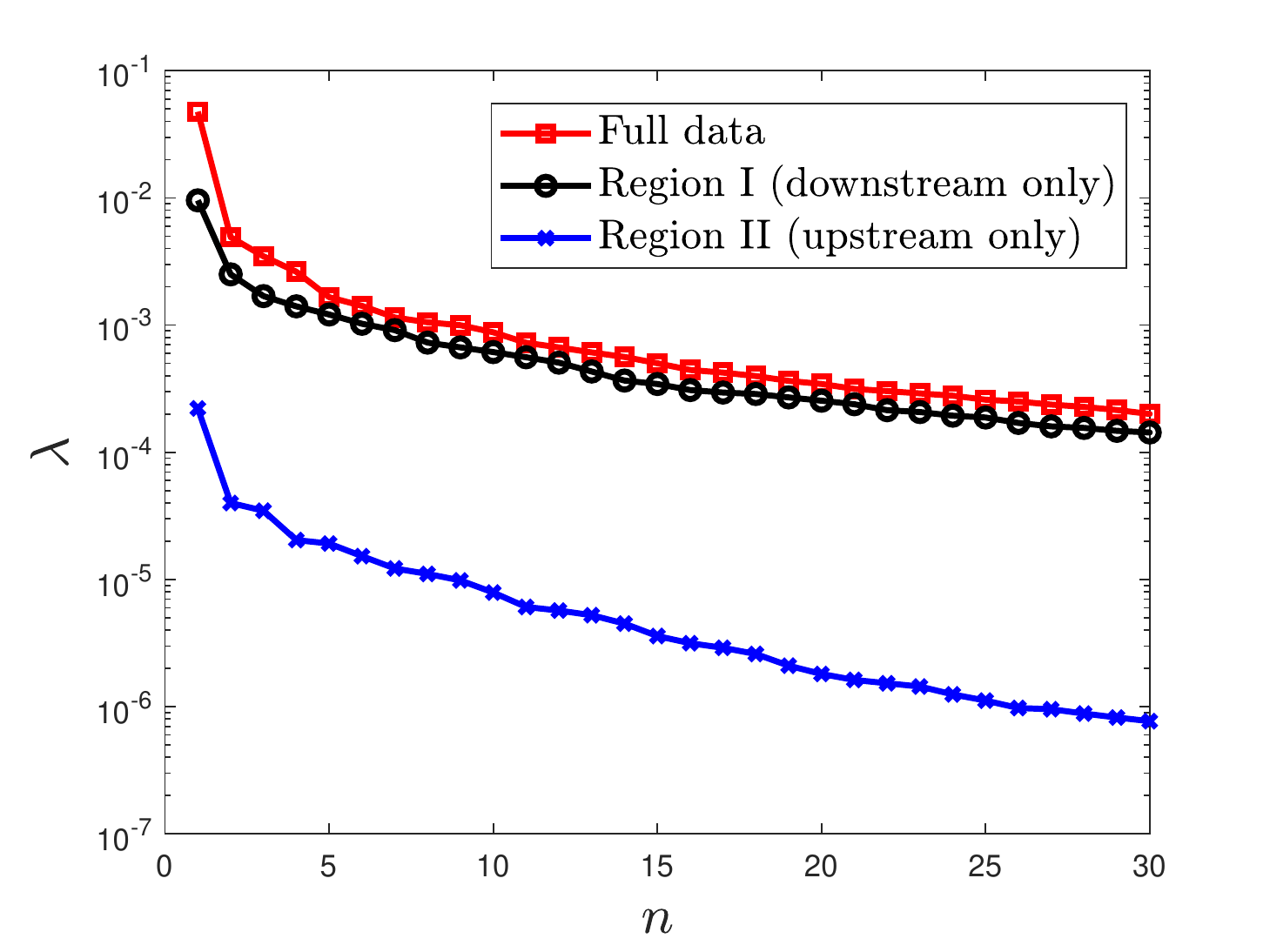}
\end{center}
   \caption{Eigenvalues of the SPOD calculation obtained from the full field and regions I/II at $St=0.03$.}
  \label{eigenvaluesoffre03and003reconstructed}
\end{figure}

The resulting eigenvalues of the SPOD modes at $St=0.03$ are shown in figure \ref{eigenvaluesoffre03and003reconstructed} for regions I and II.
The eigenvalues obtained from region I are close to the full data spectra, and approximately three orders of magnitude higher than those computed from region II, upstream of the shock.
The corresponding first SPOD modes (not shown here for brevity) show that shock wave fluctuations can be recovered from both upstream and downstream regions. 
However, the amount of energy which is present in the shock is much lower when the upstream region is selected. 
In addition, when the upstream area (region II) is considered, the spatial footprint of the upstream travelling wave becomes much thinner when compared to the downstream travelling wave. 
These results provide further evidence of the relationship between the low-frequency motion, characteristic of the shock motion, and the flow regions downstream of it.

\section{Conclusions}

This study considered the computation of linear, single-input single-output transfer functions to detect flow regions responsible for the low-frequency shock oscillations in a SBLI interaction.
The transfer functions were used to predict unsteady shock fluctuations in a range of frequencies determined via low-pass data filtering ($St<0.3$). 
It was observed that locations downstream of the shock could be used to accurately predict its fluctuations, where correlations of up to 0.6 were obtained between predicted and LES signals, even for the largest streamwise separation between input and output signals. 
The correlation magnitude strongly depends on data spanwise averaging, pointing to a predominantly 2D mechanism governing the shock unsteadiness.

Evaluation of the transfer functions in the frequency domain by means of a Hilbert transform suggests a causal relationship between flow regions downstream of the shock and its motion. 
When combined to the observed correlations, we conclude that the only causal inputs leading to high correlations with the shock are the locations surrounding and downstream of the recirculation bubble.
In these regions, the fluctuations exhibit a negative phase velocity. 
These observations shed further light on the mechanisms suggested in previous investigations \cite{Touber2009,Piponniau2009}. 
Causality was also detected between the upstream boundary layer and the shock motion, as reported in \citet{Ganapathisubramani2007}, but with a much lower correlation when compared to downstream locations.

The computation of the spectral modes from the SPOD analysis pointed to the existence of an upstream traveling wave in the leading mode at $St=0.03$. 
The Fourier transform of the leading mode confirmed the negative group velocity for frequencies up to $St\approx 0.1$ in the near-wall region. 
On the other hand, the same mode presented a positive group velocity at higher wall-normal positions. 
In addition, we demonstrated that a significant portion of the energy of the shock motion can be recovered from a limited set of measurements taken downstream of the shock.
Finally, comparison of the corresponding eigenvalues indicated that the downstream region contributes more significantly to the shock unsteadiness than the upstream region.

\section{Acknowledgments}
This work was supported by the High Performance Computing resources of the Institut du Développement et des Ressources en Informatique Scientifique under the allocation 2014-2a1877 and 2015-2a1877 awarded by Grand Équipement National de Calcul Intensif for the LES computations. A.V.G.C. aknowledges funding from CNPq (grant 310523/2017-6).

\bibliographystyle{plainnat}
\bibliography{biblio}

\end{document}